\documentclass[preprint,authoryear,11pt,fleqn]{elsarticle}
\usepackage{amssymb}
\usepackage{color}
\usepackage{amsmath}
\usepackage{graphicx}
\usepackage{graphics}
\usepackage{booktabs}
\usepackage{threeparttable}
\usepackage{appendix}
\usepackage{geometry}
\usepackage[displaymath]{lineno}
\usepackage{setspace}
\usepackage[colorlinks,linkcolor=red,anchorcolor=blue,citecolor=green]{hyperref}
\usepackage{soul}
\usepackage{subcaption}
\usepackage{listings}

\usepackage{mathpazo}

\usepackage{titlesec}
\titlespacing*{\subsection}{0pt}{*3}{*1.5}

\journal{ }

\begin{document}
\begin{spacing}{1.1}

\begin{frontmatter}

%\title{Revisiting Edie’s Generalized Definitions of Traffic Flow and Constructing Fundamental Diagrams Using Large-Scale Vehicle Trajectory Data}

\title{Constructing the fundamental diagrams of traffic flow from large-scale vehicle trajectory data}

%\author{Zhengbing He}
%\address{College of Metropolitan Transportation} 
%\address{Beijing University of Technology, Beijing, China}
%\address{Email:he.zb@hotmail.com}

%\author{Zhengbing He}
%\address{Beijing Key Laboratory of Traffic Engineering, Beijing University of Technology, Beijing, China\\ \vspace{0.1in} Email: he.zb@hotmail.com}

\author{Zhengbing He\corref{cor1}}
\ead{he.zb@hotmail.com}

\author{Cathy Wu}

\address[rvt1]{Laboratory for Information and Decision Systems (LIDS)
Massachusetts Institute of Technology, Cambridge, United States, MA 02139}
\cortext[cor1]{Corresponding author}

\begin{abstract}
For decades, researchers and practitioners typically measure macroscopic traffic flow variables, i.e., density, flow, and speed, using time or space cuts, and then construct the fundamental diagrams of traffic flow.
With the advent of large-scale vehicle trajectory datasets, often capturing 100\% of vehicle dynamics, Edie’s generalized definitions have become widely recognized as the most appropriate framework for measuring these variables.
However, while Edie’s formulation explicitly requires the traffic state within the measurement region to be both stationary and homogeneous, there is little guidance on how to systematically identify such qualified regions and construct the corresponding fundamental diagrams.
To address this gap, this paper proposes an Edie’s definition-based method for measuring traffic variables and constructing the fundamental diagrams of traffic flow by automatically identifying stationary traffic states using parallelogram-shaped aggregation regions.
An open-source tool is developed and released to support both researchers and practitioners.
From now on, we have an automated tool that can generate fundamental diagrams directly from any large-scale time-space diagram of vehicle trajectories, either \textbf{collected from the real world} or \textbf{generated by simulation} (such as testing car-following models).

\vspace{7mm}

\noindent \textit{Code}: \url{https://github.com/gotrafficgo/construct_fundamental_diagram} 

\vspace{1mm}

\noindent \textit{Animation}: \url{https://m.youtube.com/watch?v=lJVYIVtsLso}

\end{abstract}

\begin{keyword}
Traffic flow theory \sep fundamental diagram \sep trajectory data \sep traffic dynamics \sep simulation

\end{keyword}

\end{frontmatter}

\newpage

%\tableofcontents

%\linenumbers


\newpage

%%%%%%%%%%%%%%%%%%%%%%%%%%%%%%%%%%%%%%%%%%%%%%%%%%%%%%%%%%%%%%%%%%%%%%%%%%%%%%%%%%%%%%
%%%%%%%%%%%%%%%%%%%%%%%%%%%%%%%%%%%%%%%%%%%%%%%%%%%%%%%%%%%%%%%%%%%%%%%%%%%%%%%%%%%%%%
\section{Introduction}

Macroscopic traffic flow variables, i.e., typically density, flow, and speed, are key indicators of traffic states and infrastructure performance. Pairwise combinations of these variables form the foundational relationships in traffic flow theory.
Abstracting from the inherent heterogeneity and non-stationarity of real-world traffic, their empirical characteristics are often captured using an equilibrium relationship, commonly known as the fundamental relationship, which is typically represented by three diagrams: the fundamental diagrams of flow-density, speed-density, and speed-flow. 
Macroscopic traffic flow variables and the fundamental diagram are among the oldest and most foundational concepts in the field of traffic flow theory and transportation engineering \citep{TRB2008Greenshields75}.

Although it is widely recognized that macroscopic traffic flow variables should be measured under homogeneous and stationary traffic conditions \citep{Laval2011}, this is often not achievable in practice. 
Due to the long-standing reliance on loop detectors for data collection, such ideal conditions are rarely captured when constructing fundamental diagrams. 
This has led to a situation where, even today — despite the availability of large-scale, high-resolution vehicle trajectory data (e.g., the NGSIM dataset \citep{NGSIM2006}, the Zen Traffic Dataset \citep{Dahiyal2020}, the highD dataset \citep{Kruber2019}, I-24 MOTION \citep{Gloudemans2023}, and the MiTra dataset \citep{Chaudhari2025}, as well as simulated trajectory data from car-following models and microscopic traffic simulation tools \citep{ZHENG2017136,8434122,Xie02012020,11015830,Shubo2025}) — many researchers still extract macroscopic traffic variables using virtual loop detectors or rectangular time-space cells for data aggregation \citep{Li2020a}.

\begin{figure}
    \centering
    \includegraphics[width=6in]{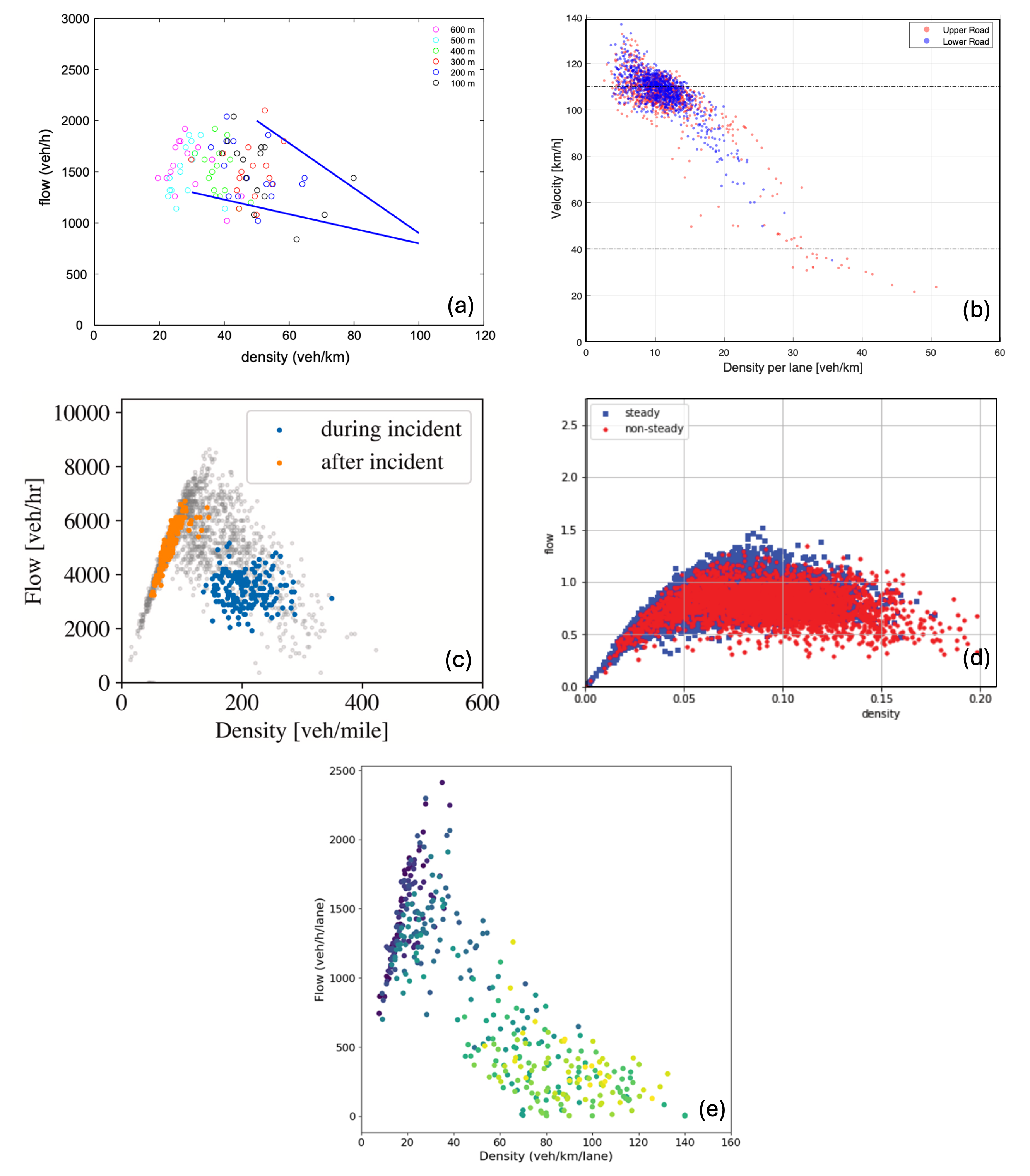}
    \caption{The existing empirical fundamental diagrams constructed from large-scale, high-resolution vehicle trajectory data.
    (a) NGSIM \citep{He2015a} (b) highD \citep{Kruber2019} (c) I-24 MOTION \citep{Gloudemans2023} (d) ZenTraffic \citep{Dahiyal2020} (e) MiTra \citep{Chaudhari2025}}
    \label{fig:Empirical}
\end{figure}

For example, \cite{He2015a} constructed the fundamental relationship by placing virtual loop detectors within the NGSIM trajectory dataset; Figure \ref{fig:Empirical}(a). 
When publishing the highD dataset, \cite{Kruber2019} measured macroscopic variables based on time and space cut, respectively; Figure \ref{fig:Empirical}(b) 
As for the Zen Traffic dataset, \cite{Dahiyal2020} divided the time-space plane into rectangular cells and applied Edie’s generalized definitions to compute flow, speed, and density \citep{Edie1965}; Figure \ref{fig:Empirical}(d). 
Similarly, the I-24 MOTION project adopted the same rectangular cell-based aggregation approach to construct fundamental diagrams \citep{Gloudemans2023}; Figure \ref{fig:Empirical}(c).
Even for the most recent trajectory dataset, the MiTra dataset, which was collected on the A50 urban freeway in Milan, Italy and released nearly simultaneously with this paper, the fundamental diagram was constructed using rectangular cell-based Edie’s definitions \citep{Chaudhari2025}; Figure \ref{fig:Empirical}(e).
Clearly, none of these practices, including those adopted by researchers who published high-fidelity trajectory datasets, effectively capture homogeneous and stationary traffic conditions, and they substantially underutilize the richness and granularity of modern trajectory data.

To maximize the likelihood of capturing stationary traffic conditions within the selected aggregation area, \cite{Laval2011} developed an automated tool called ``Trajectory Explorer," which aggregates traffic data using a series of small parallelograms aligned with vehicle platoons of fewer than 10 vehicles.
By using this tool, a new shape of the hysteresis loop in the fundamental diagram was uncovered, underscoring the importance of explicitly accounting for stationarity in measuring traffic variables.
Although this tool has demonstrated effective, it is not readily applicable to constructing fundamental diagrams from large-scale spatiotemporal trajectory datasets.

In summary, although large-scale spatiotemporal trajectory datasets are now widely available, existing methods remain inadequate for effectively constructing fundamental diagrams. To fill this gap, this paper proposes a new tool that can automatically generate the fundamental diagram directly from a large-scale trajectory dataset, significantly outperforming the commonly used approaches.
To promote easy use and reproducibility, the code has been released in \url{https://github.com/gotrafficgo/construct_fundamental_diagram}.

%%%%%%%%%%%%%%%%%%%%%%%%%%%%%%%%%%%%%%%%%%%%%%%%%%%%%%%%%%%%%%%%%%%%%%%%%%%%%%%%%%%%%%
%%%%%%%%%%%%%%%%%%%%%%%%%%%%%%%%%%%%%%%%%%%%%%%%%%%%%%%%%%%%%%%%%%%%%%%%%%%%%%%%%%%%%%
\section{Background}

\subsection{Three types of time-space intervals for measuring traffic flow}

\begin{figure}
    \centering
    \includegraphics[width=2.7in]{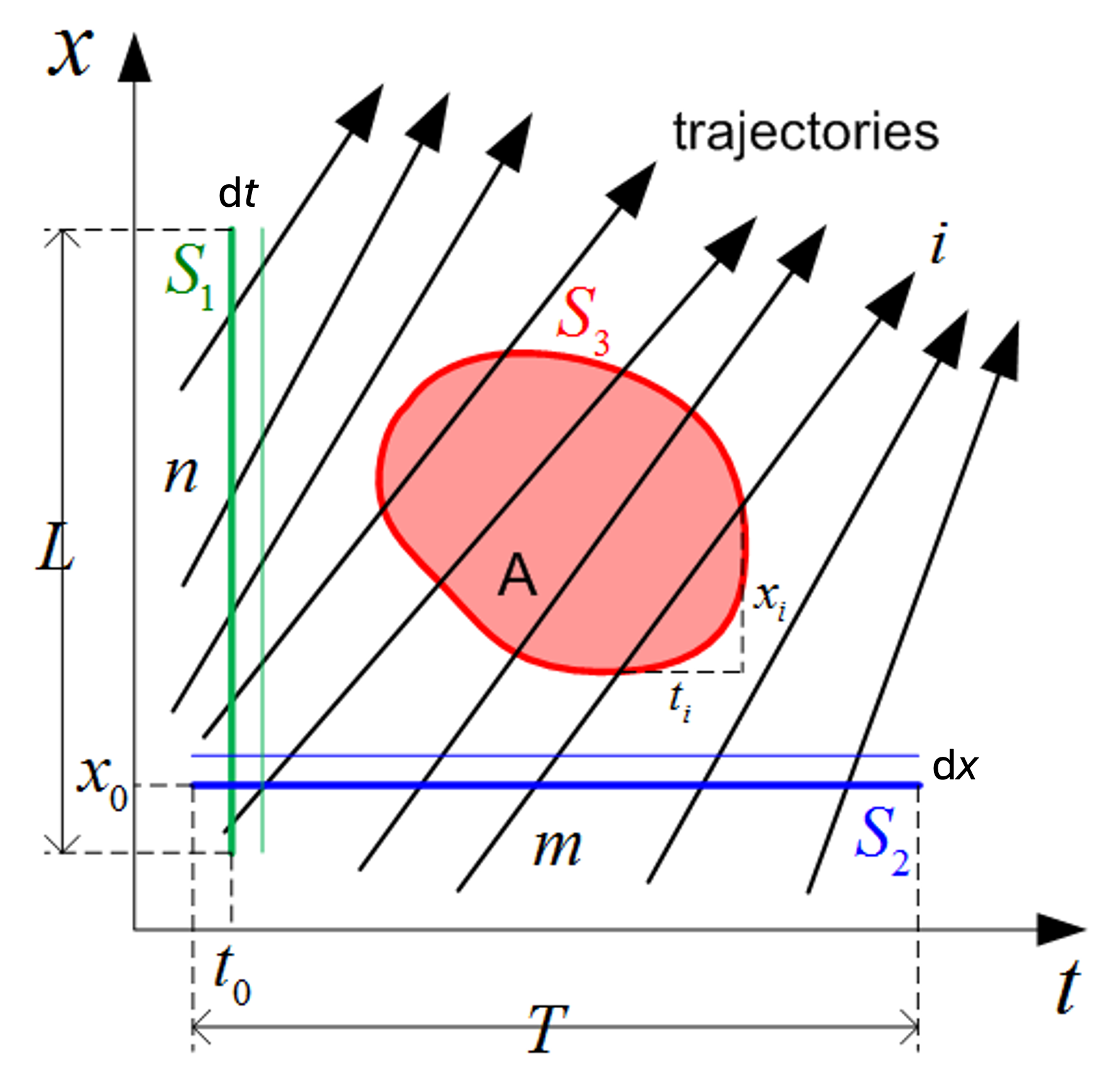}
    \caption{Three types of time-space intervals for measuring traffic flow.}
    \label{fig:Measurement}
\end{figure}

In general, there are three types of time-space intervals for measuring traffic flow variables (Figure \ref{fig:Measurement}):
\begin{itemize}
    \item \textbf{S1}: Fixed location with continuous time, typically obtained from aerial photography.
    \item \textbf{S2}: Fixed time with continuous location, typically measured using loop detectors.
    \item \textbf{S3}: Continuous time and continuous space, typically captured by roadside or drone-mounted cameras.
\end{itemize}

\subsection{Calculations of macroscopic traffic flow variables}

The most commonly used macroscopic traffic flow variables are density, flow, and speed.
Let $n$ and $m$ denote the number of the vehicles within segment $L$ at time $t_0$ and the number of the vehicles passing cross-section $x_0$ during time period $T$ (Figure \ref{fig:Measurement}).
Let $x_i$ and $t_i$ be the travel distance and travel time, respectively, of the $i$-th vehicle within the time-space region $A$. Then, the total travel distance and total travel time of all vehicles within region $A$ are denoted by $d(A)$ and $t(A)$, respectively.
Let $|A|$ be the area of region A.

According to the three types of time-space intervals, the calculations of the three macroscopic traffic flow variables are as follows.
Note that the equations for \textbf{S3} are formulated according to Edie’s generalized definitions of flow, speed and density \citep{Edie1965}.

\subsubsection{Density $k$ (veh/km)}

Density is defined as \textit{vehicle number per distance unit at a time slice}.
Therefore, according to the definition of density,  
\begin{equation}
	\textbf{S1 - Density}: \ \ \ \  k(x, t_0, L) = \frac{n}{L}.
\end{equation}
By expanding \textbf{S1} interval to an area, 
\begin{equation}\label{equ:density_s3}
	\textbf{S3 - Density}: \ \ \ \  k(x, t, A) = \frac{n}{L} \approx \frac{t(A)}{|A|}.
\end{equation}
By applying Equation \ref{equ:density_s3} for \textbf{S3} to \textbf{S1} area, 
\begin{equation}\label{equ:density_s2}
	\textbf{S2 - Density}: \ \ \ \  k(x_0, t, T) = \frac{\sum_m \frac{\mathrm{d} x}{v_i}}{T\mathrm{d}x} = \frac{\sum_m\frac{1}{v_i}}{T}.
\end{equation}

\subsubsection{Flow $q$ (veh/h)}

Flow is defined as \textit{vehicle number passing a cross-section per time unit}.
Therefore, according to the definition of flow,  
\begin{equation}\label{equ:flow_s2}
	\textbf{S2 - Flow}: \ \ \ \  q(x_0, t, T) = \frac{m}{T}.
\end{equation}
By expanding \textbf{S2} interval to an area, 
\begin{equation}\label{equ:flow_s3}
	\textbf{S3 - Flow}: \ \ \ \  q(x, t, A) = \frac{m\mathrm{d}x}{T\mathrm{d}x} \approx \frac{d(A)}{|A|}.
\end{equation}
By applying Equation \ref{equ:flow_s3} for \textbf{S3} to \textbf{S1} area, 
\begin{equation}
	\textbf{S1 - Flow}: \ \ \ \  q(x, t_0, L) = \frac{\sum_n v_i \mathrm{d}t}{L\mathrm{d}t} = \frac{\sum_n v_i}{L}.
\end{equation}

\subsubsection{Speed $v$ (km/h)}

(Space-mean) speed is defined as \textit{the quotient of the flow and the density}
Therefore, according to the definition of speed,  
\begin{equation}
	\textbf{S1 - Speed}: \ \ \ \  v(x, t_0, L) = \frac{\frac{\sum_n v_i}{L}}{\frac{n}{L}} = \frac{1}{n}\sum_n v_i.
\end{equation}
%%%
\begin{equation}\label{equ:speed_s2}
	\textbf{S2 - Speed}: \ \ \ \  v(x_0, t, T) = \frac{\frac{m}{T}}{\frac{\sum_m\frac{1}{v_i}}{T}} = \frac{m}{\sum_m \frac{1}{v_i}}.
\end{equation}
%%%
\begin{equation}\label{equ:speed_s3}
	\textbf{S3 - Speed}: \ \ \ \  v(x, t, A) = \frac{d(A)}{t(A)}
\end{equation}

%%%%%%%%%%%%%%%%%%%%%%%%%%%%%%%%%%%%%%%%%%%%%%%%%%%%%%%%%%%%%%%%%%%%%%%%%%%%%%%%%%%%%%
%%%%%%%%%%%%%%%%%%%%%%%%%%%%%%%%%%%%%%%%%%%%%%%%%%%%%%%%%%%%%%%%%%%%%%%%%%%%%%%%%%%%%%
\section{Methodology}\label{sec:Methodology}

\subsection{Parallelogram-shaped aggregation region}

Given a universally consistent wave speed\footnote{Stop-and-go waves propagate against the traffic direction at a speed of 10$\sim$20 km/h, i.e., the slope of the waves in the time-space diagram or backward-moving speed of the queues \citep{Mauch2002a,Laval2010,Zheng2011a,Jiang2014,He2015a,He2024}.} $\omega$ and a traffic speed $v^*$ of interest, we construct a parallelogram-shaped aggregation region to extract relevant trajectory data points. As shown in previous studies \citep{Laval2011,4914846,Ahn2013,He2019,dts-0024-0001}, it is more effective to use regions whose two longer sides align with the wave speed.
The slopes of the remaining two sides are set to be the traffic speed $v^*$.

\subsection{Score for stationary traffic}

With a parallelogram-shaped aggregation region and the trajectory points inside, we then evaluate whether the collected data is sufficiently stationary using the following two metrics: the coefficient of variation (CV) and the normalized absolute error (NAE). 
%%%
The CV measures the relative variability of speed observations:
\begin{equation}
	\text{CV} = \frac{\sigma_v}{\mu_v}
\end{equation}
where $\sigma_v$ and $\mu_v$ are the standard deviation and mean of the observed speeds of the data points, respectively.
%%%
The NAE evaluates the average deviation of observed speeds $v_i$ from the given speed $v^*$, normalized by the maximum magnitude among each observed speed, the reference speed, and a small constant $\varepsilon$ to avoid division by zero:
\begin{equation}
	\text{NAE} = \frac{1}{n} \sum_{i=1}^{n} \frac{ \left| v_i - v^* \right| }{ \max \left( |v_i|, |v^*|, \varepsilon \right) }
\end{equation}
where $n$ is the number of trajectory points inside the aggregation region, and $\varepsilon$ is a small constant (e.g., $10^{-3}$).
%%%
Eventually, we use a weighted metric to jointly consider both CV and NAE:
\begin{equation}
	\text{Score} = w_\text{CV} \cdot \text{CV} + w_\text{NAE} \cdot \text{NAE}
\end{equation}
where $w_\text{CV}$ and $w_\text{NAE}$ are the weights of the CV and NAE, respectively.
Note that traffic models usually require homogeneous traffic, meaning all vehicles are essentially identical. Here, we ignore this requirement; however, one could further refine the data by adding additional conditions, since vehicle length information is also included in some of trajectory datasets.

\subsection{Measurement of traffic variables from the data in a parallelogram}

\begin{figure}
    \centering
    \includegraphics[width=2.5in]{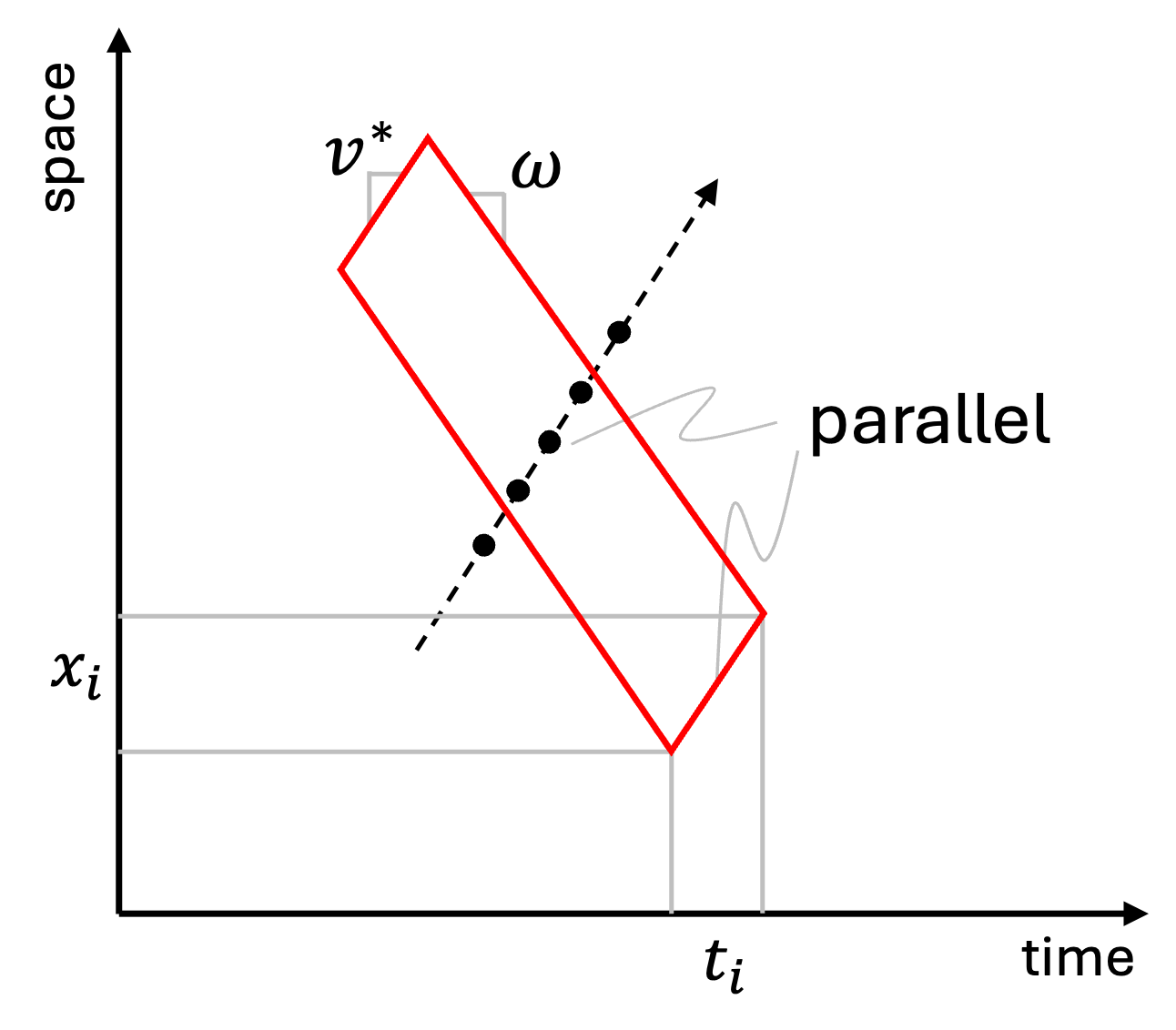}
    \caption{Parallelogram and trajectory points inside.}
    \label{fig:Parallelogram}
\end{figure}

Assume that we are able to align the speeds of all trajectory points within a parallelogram to the target speed $v^*$, by selecting the parallelogram with small ``Score".
Since the slope of the parallelogram’s shorter sides is equal to $v^*$, any trajectory passing through the region will be approximately parallel to these sides. Consequently, the travel distance $x_i$ and travel time $t_i$ of the $i$-th vehicle can be approximated by the projections of the shorter sides onto the spatial and temporal axes, respectively (Figure \ref{fig:Parallelogram}).
Therefore, the calculation using Edie's definition (i.e., Equations \ref{equ:density_s3}, \ref{equ:flow_s3}, \ref{equ:speed_s3}) can be re-written as follows.
\begin{equation}\label{equ:density_s3+}
	\textbf{S3 - Density}: \ \ \ \  k(x,t,A) \approx N\frac{x_i}{|A|}	
\end{equation}
\begin{equation}\label{equ:flow_s3+}
	\textbf{S3 - Flow}: \ \ \ \  q(x,t,A) \approx N\frac{t_i}{|A|}	
\end{equation}
\begin{equation}\label{equ:speed_s3+}
	\textbf{S3 - Speed}: \ \ \ \  v(x,t,A) \approx \frac{x_i}{t_i} = v^*
\end{equation}
Since the size of the parallelogram is predefined and identical, the values of $x_i$ and $t_i$ are the same for all vehicles within the region. Therefore, both the density and the flow depend solely on the number of vehicles contained within the parallelogram.

\subsection{Generation of all parallelograms}

We span the given speed $v^*$ from the jam speed (zero) to the free-flow speed using a fixed speed step.
For a given speed $v^*$, we typically generate tens of thousands of parallelograms to capture as many qualified regions as possible. There are two practical tips that greatly influence performance.
\begin{itemize}
\item Instead of placing parallelograms completely at random, significant time can be saved by centering them on data points whose speeds ($v'$) are very close to $v^*$. This approach initially filters out points with speeds far from $v^*$, which essentially have no chance of qualifying.
In the code, we primarily set an expected number of candidate parallelograms, and adjust the tolerance of the speed difference between $v^*$ and $v'$ until either a sufficient number of parallelograms is found or the maximum allowed tolerance is reached.

\item When placing parallelograms, new ones should not overlap with any existing parallelograms, regardless of the associated given speed. This ensures that no parallelograms overlap in the final result.

\end{itemize}

\subsection{Selection of parallelograms}

We sort all parallelograms according to their scores and select the top $M$ as the final set.
Then, the fundamental diagram could be constructed according to Equations \ref{equ:density_s3+}, \ref{equ:flow_s3+}, and \ref{equ:speed_s3+}.

%%%%%%%%%%%%%%%%%%%%%%%%%%%%%%%%%%%%%%%%%%%%%%%%%%%%%%%%%%%%%%%%%%%%%%%%%%%%%%%%%%%%%%
%%%%%%%%%%%%%%%%%%%%%%%%%%%%%%%%%%%%%%%%%%%%%%%%%%%%%%%%%%%%%%%%%%%%%%%%%%%%%%%%%%%%%%
\section{Results}\label{sec:Results}

To demonstrate the proposed method, we employ two well-known high-fidelity vehicle trajectory datasets: the ZenTraffic dataset from Japan and the NGSIM dataset from the United States.
\begin{itemize}
	\item The employed ZenTraffic data was collected on a 2 km stretch of the Hanshin Expressway Route 11 (Ikeda Line) near the Tsukamoto junction in Osaka, Japan. Using 38 high-mounted rear-facing cameras and advanced image recognition techniques, vehicle trajectories were captured at 0.1-second intervals. The dataset includes five 1-hour time windows (i.e., F001$\sim$F005), each recording around 3,000 to 4,000 vehicle trajectories with high temporal and spatial resolution.

	\item The employed NGSIM data was recorded on a 640 m section of US-101 (Hollywood Freeway) in Los Angeles, California. The data focuses on a congested merging area with five lanes in the southbound direction and was collected at 0.1-second intervals using a network of synchronized video cameras. There are three data segments, covering the time periods 7:50-8:05 AM, 8:05-8:20 AM, and 8:20-8:35 AM.
\end{itemize}
To improve processing speed, the data were previously aggregated at a 1-second interval by averaging the values within each second.

The slopes of the long sides of the parallelograms are set to the wave speeds, which are -15 km/h for the ZenTraffic dataset and -16 km/h for the NGSIM dataset, respectively. The wave speed could be visually measured from the time-space diagram of trajectories, which is quite straightforward.

Each parallelogram has long sides of length 100 in the space-time plane, with a perpendicular distance (i.e., height) of 5 between them. Thus, the area of each parallelogram is 500.
In addition, $w_\text{CV} = 0.5$ and $w_\text{NAE}=0.5$.

Note that due to space limitations, we only present a limited number of settings and results. Since the code has been open-sourced, users can easily modify the settings and apply their own datasets, provided that the trajectory data is preprocessed into the following format, where each row corresponds to a trajectory data point:
\begin{lstlisting}[basicstyle=\ttfamily\small, frame=single, caption={Required format of trajectory data}]
Column 1: Vehicle ID
Column 2: Time (s)
Column 3: Location (m)
Column 4: Speed (km/h)
\end{lstlisting}

The baseline is the virtual loop detector method, which places a virtual detector every 100\,m, aggregates the trajectory data passing each location every 30\,s (i.e., by averaging), and computes the traffic flow variables based on Equations~\ref{equ:density_s2}, \ref{equ:flow_s2}, and~\ref{equ:speed_s2}.

\begin{figure}[htbp]
    \centering
    \begin{subfigure}{0.45\textwidth}
        \includegraphics[width=\linewidth]{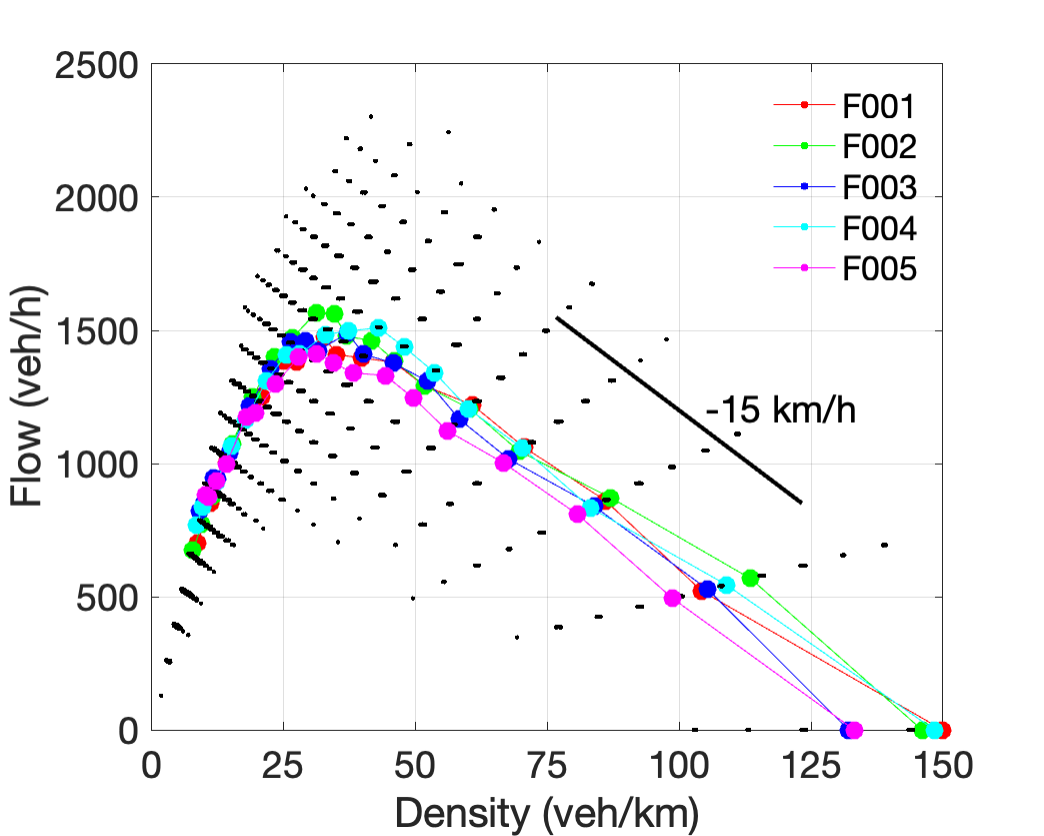}
        \caption{ZenTraffic (TRJ 11, Lane 1): The proposed one}
    \end{subfigure}
    \begin{subfigure}{0.45\textwidth}
        \includegraphics[width=\linewidth]{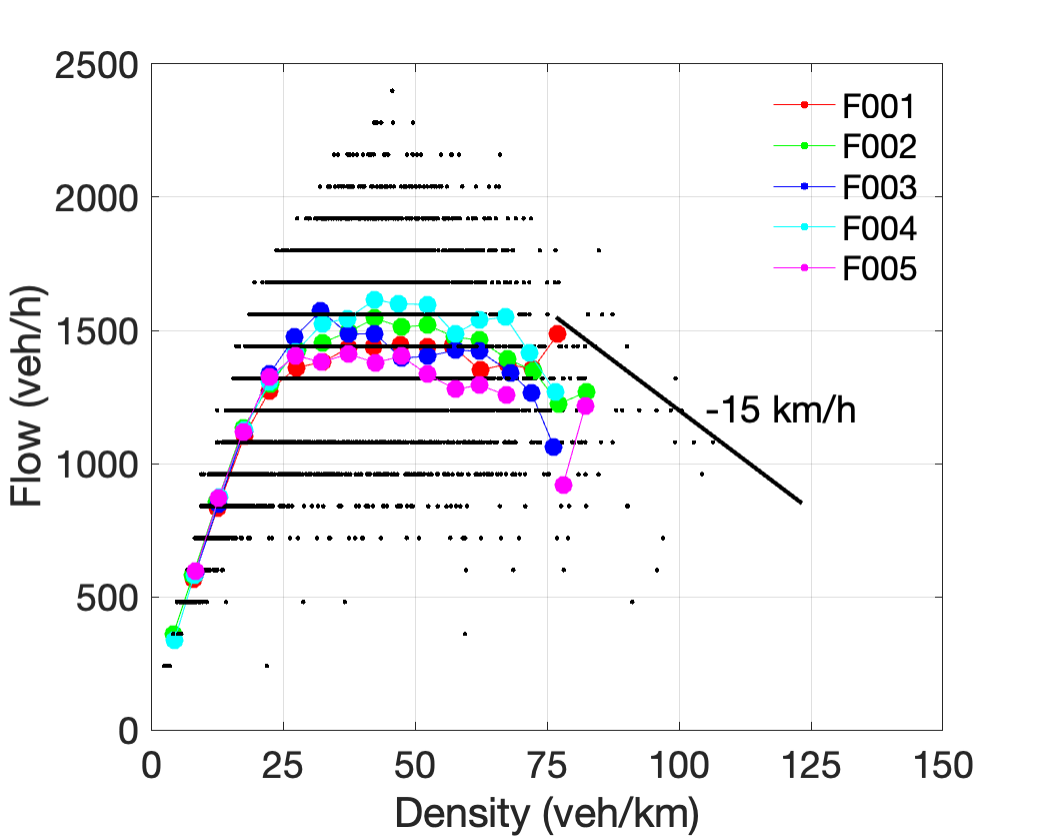}
        \caption{ZenTraffic (TRJ 11, Lane 1): Virtual loop}
    \end{subfigure}
    
    \begin{subfigure}{0.45\textwidth}
        \includegraphics[width=\linewidth]{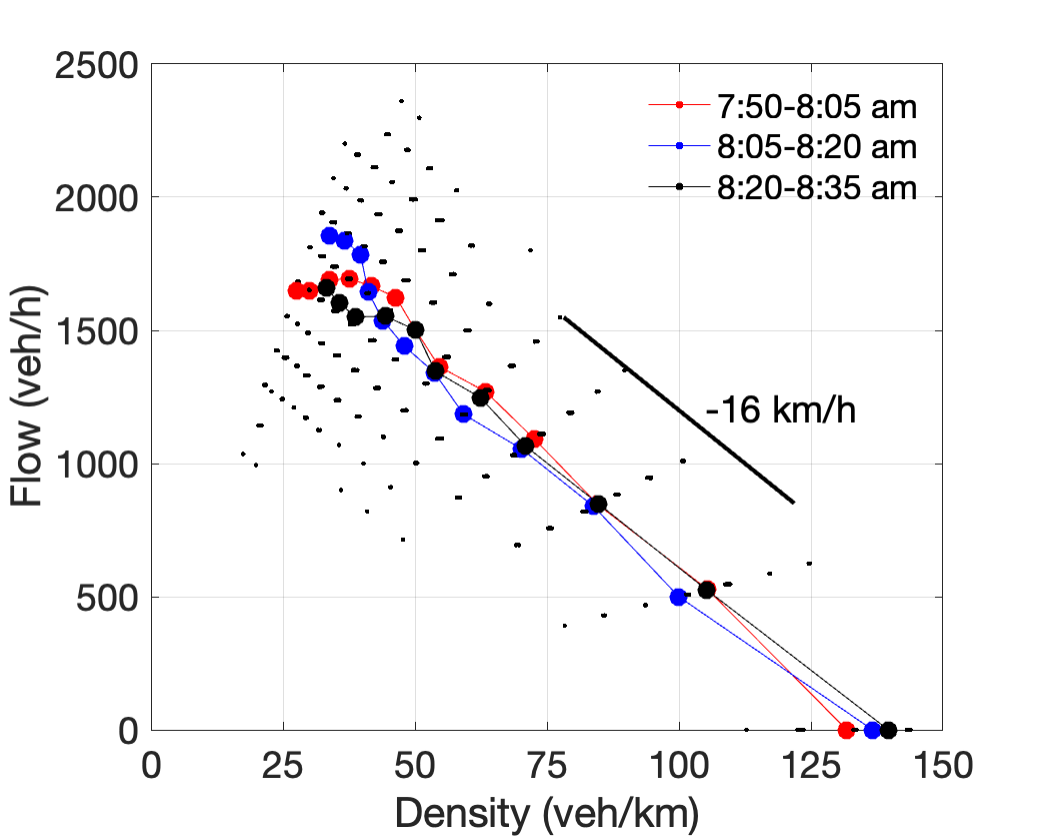}
        \caption{NGSIM (US101, Lane 1): The proposed one}
    \end{subfigure}
    \begin{subfigure}{0.45\textwidth}
        \includegraphics[width=\linewidth]{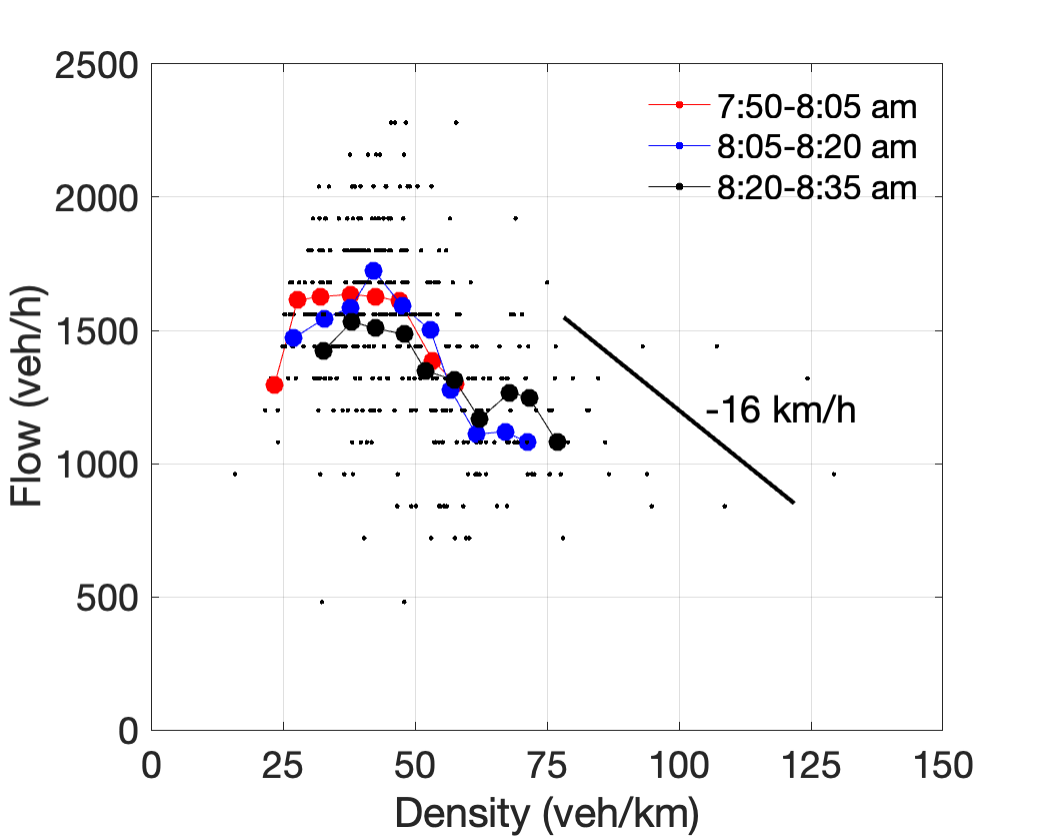}
        \caption{NGSIM (US101, Lane 1): Virtual loop}
    \end{subfigure}
    \caption{Empirical fundamental diagrams constructed with the proposed method and the traditional virtual-loop-detector method.}
    \label{fig:FD}
\end{figure}

Figure~\ref{fig:FD} presents the results. The fundamental diagrams constructed using the proposed approach clearly exhibit the congestion branch, and the wave speed exhibited on the fundamental diagrams aligns well with both the preset value and empirical observations. 
In contrast, the traditional virtual loop detector-based method fails to capture the congestion branch, as also observed in the other dataset shown in Figure~\ref{fig:Empirical}.

Note that while the black scatters in the background of the fundamental diagrams may appear similar across methods, their underlying interpretations differ significantly. In the loop detector-based method, these scatters are merely the result of periodic aggregation and do not carry specific physical meaning. In contrast, in the proposed approach, each black scatter corresponds to a small parallelogram region characterized by a locally quasi-stationary traffic state. Given that perfectly stationary states are rarely observed in real-world traffic, we extract such regions individually and assemble them to construct a fundamental diagram that better approximates stationary conditions.

\begin{figure}[htbp]
    \centering
    
    \begin{subfigure}{0.45\textwidth}
        \includegraphics[width=\linewidth]{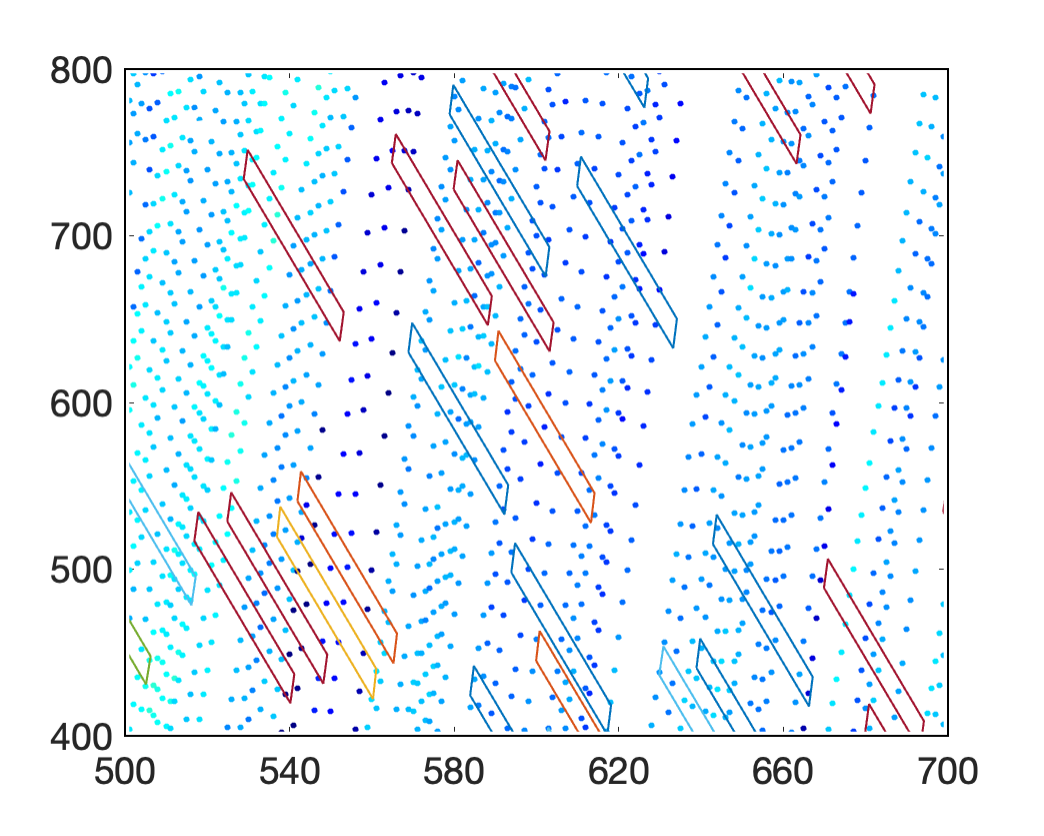}
        \caption{Zoom-in of Region A}
    \end{subfigure}
    \hspace{10mm}  % 或更小如 2mm，让图更近
    \begin{subfigure}{0.45\textwidth}
        \includegraphics[width=\linewidth]{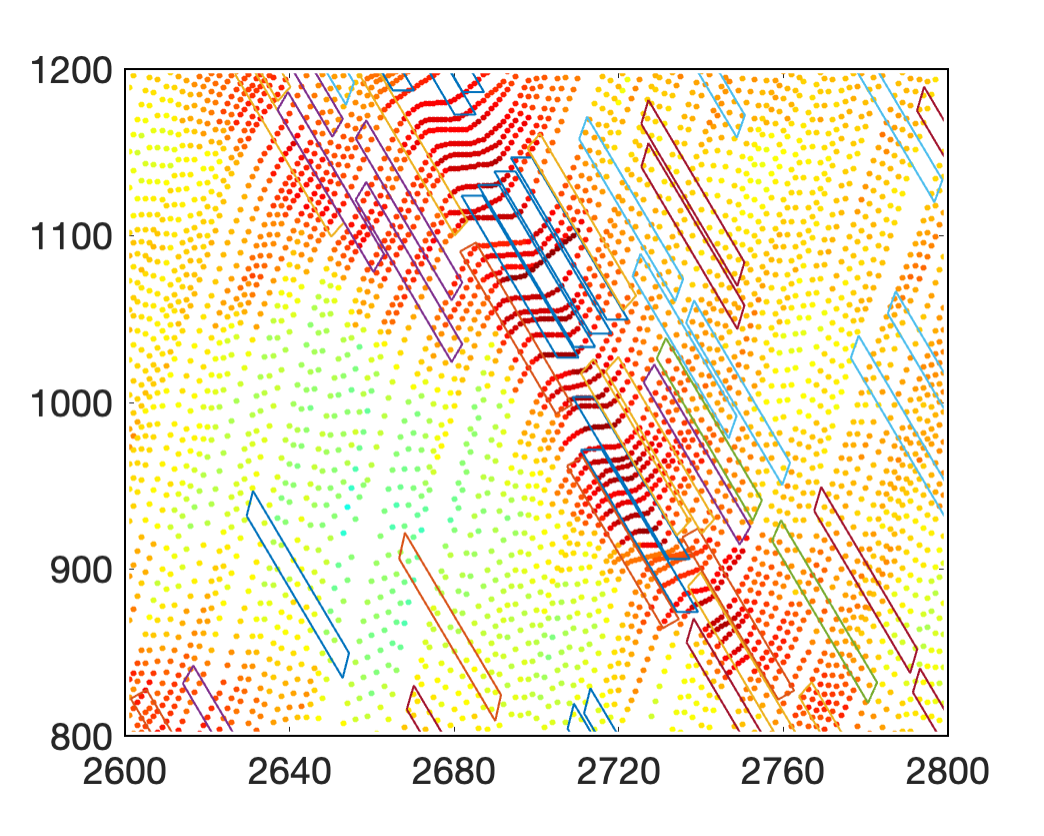}
        \caption{Zoom-in of Region B}
    \end{subfigure}

    \vspace{4mm} 

    \begin{subfigure}{\textwidth}
        \includegraphics[width=\linewidth]{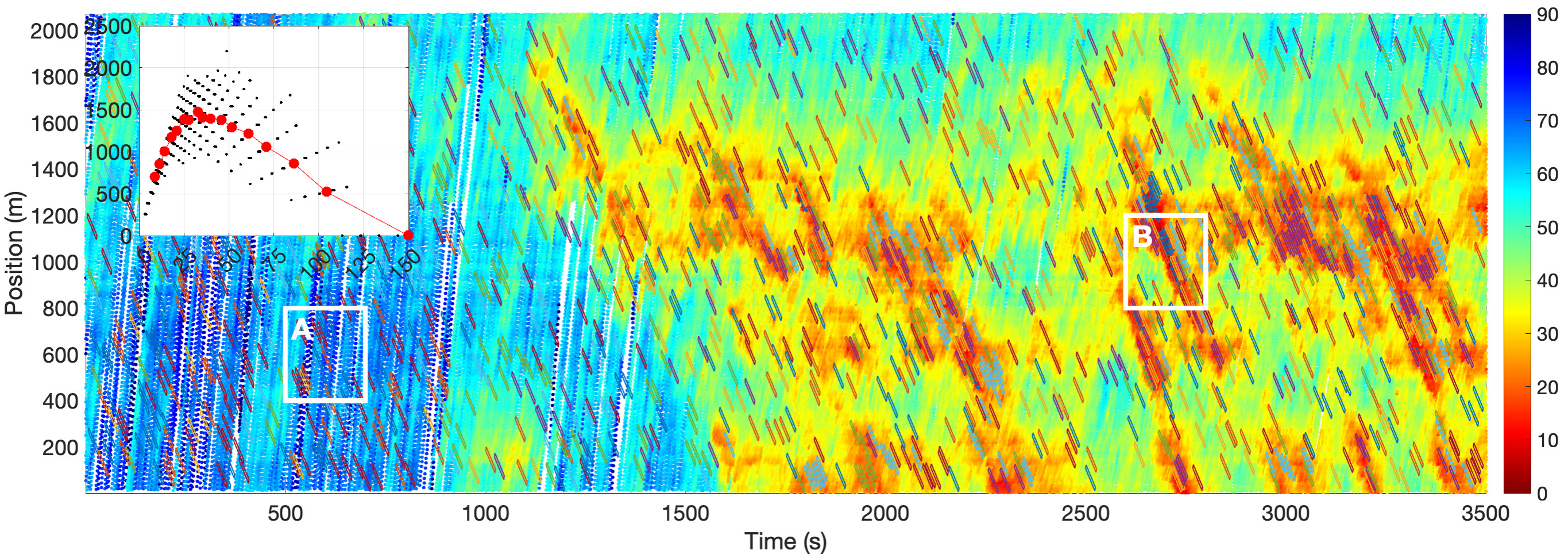}
        \caption{The whole diagrams.}
    \end{subfigure}

    \caption{The time-space diagram of trajectories and the generated parallelograms (The ZenTraffic data (TRJ 11: Lane 1, F001)).}
    \label{fig:TXPlot}
\end{figure}

Figure~\ref{fig:TXPlot} provides a detailed illustration of the data collection process employed in the proposed method.

\begin{figure}[htbp]
    \centering
    \begin{subfigure}{0.32\textwidth}
    	\includegraphics[width=\linewidth]{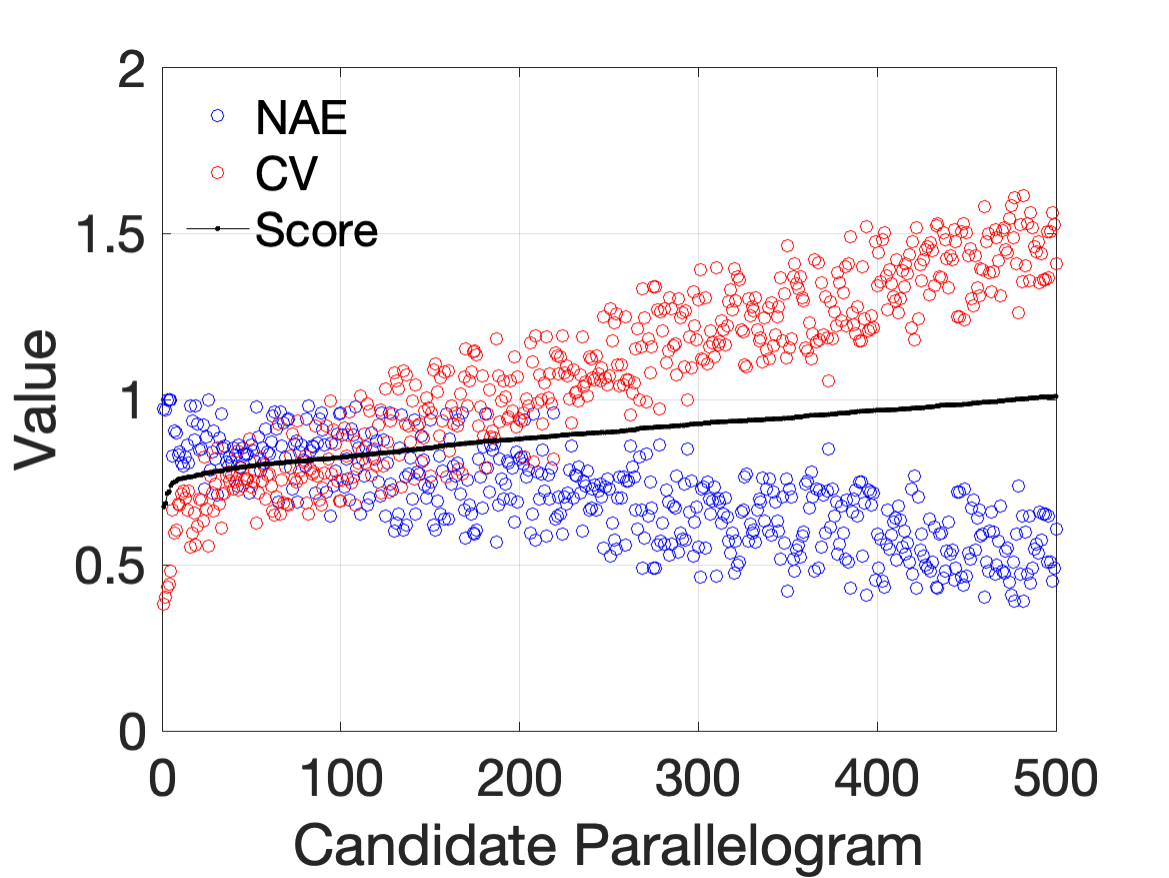}
        \caption{5 km/h}
    \end{subfigure}
    \begin{subfigure}{0.32\textwidth}
        \includegraphics[width=\linewidth]{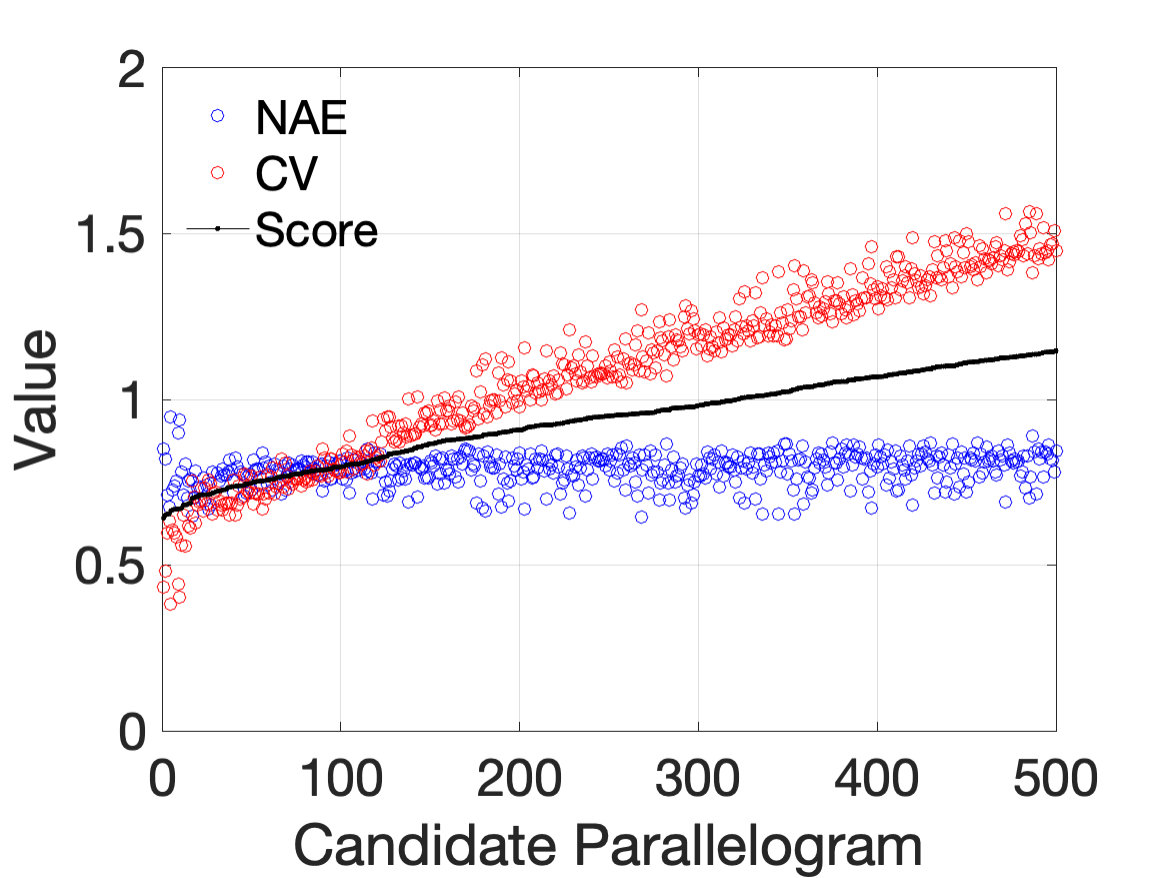}
        \caption{10 km/h}
    \end{subfigure}
    \begin{subfigure}{0.32\textwidth}
        \includegraphics[width=\linewidth]{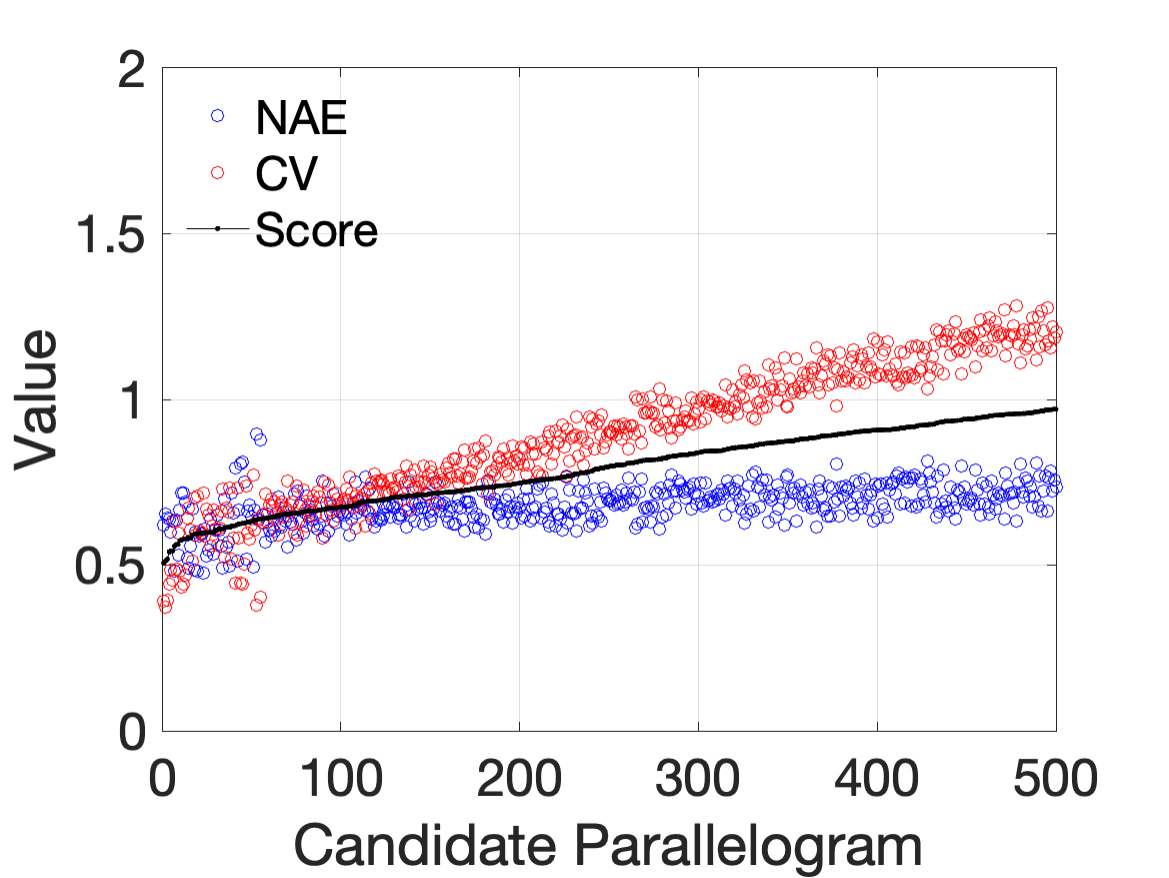}
        \caption{20 km/h}
    \end{subfigure}
    
    \begin{subfigure}{0.32\textwidth}
        \includegraphics[width=\linewidth]{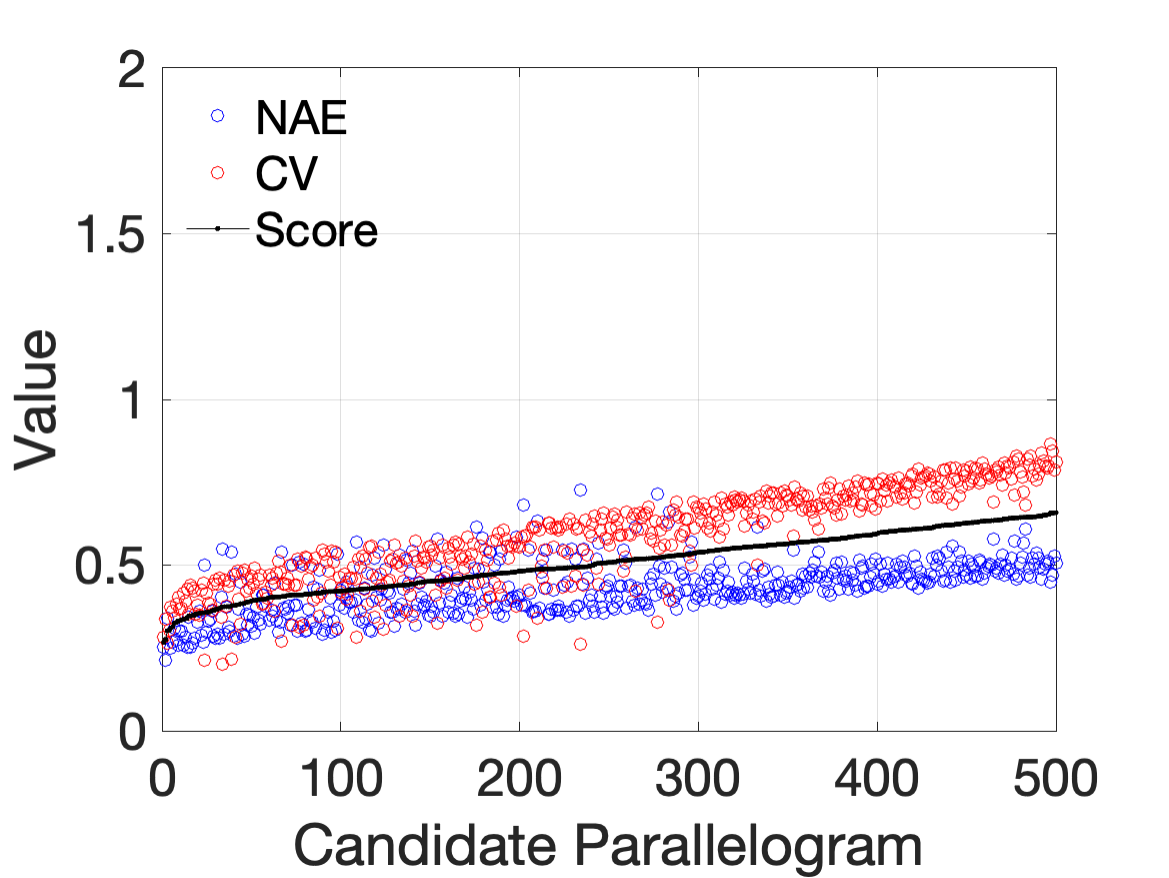}
        \caption{30 km/h}
    \end{subfigure}
    \begin{subfigure}{0.32\textwidth}
        \includegraphics[width=\linewidth]{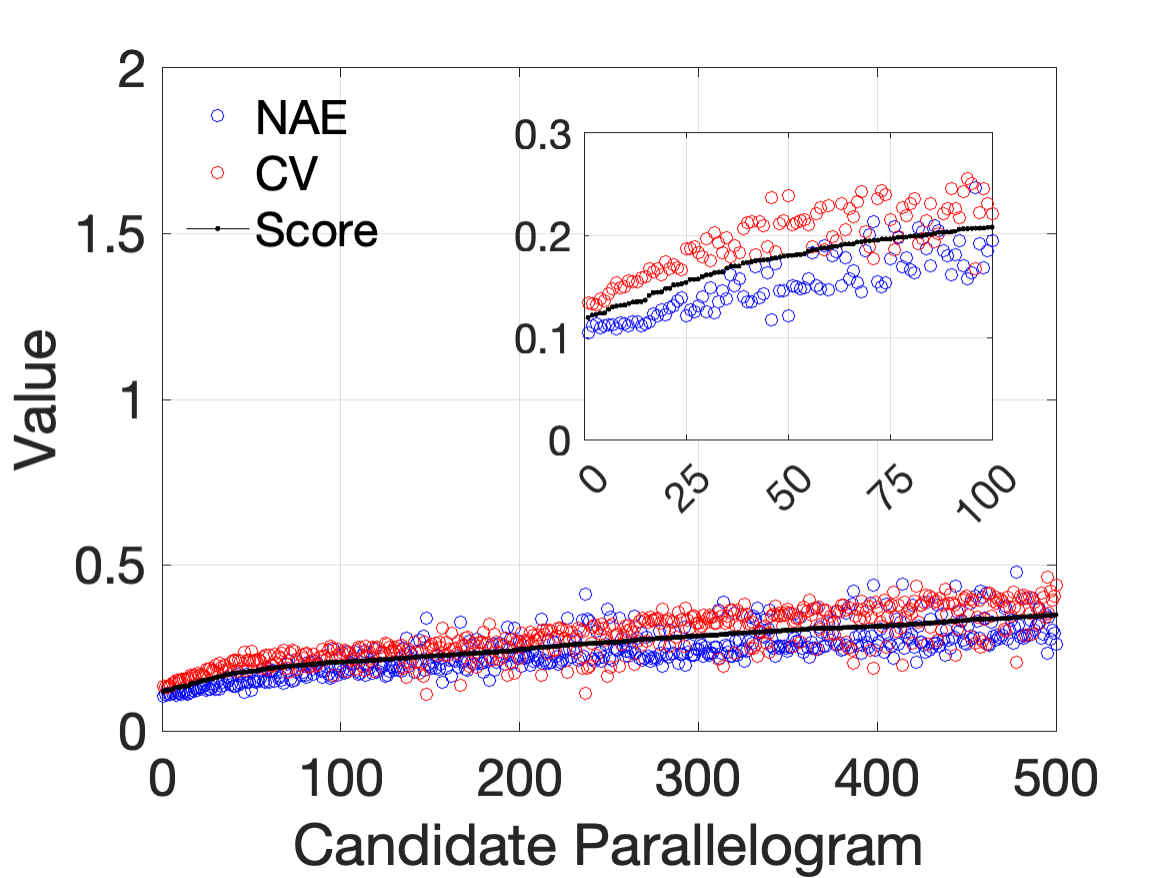}
        \caption{40 km/h}
    \end{subfigure}
    \begin{subfigure}{0.32\textwidth}
        \includegraphics[width=\linewidth]{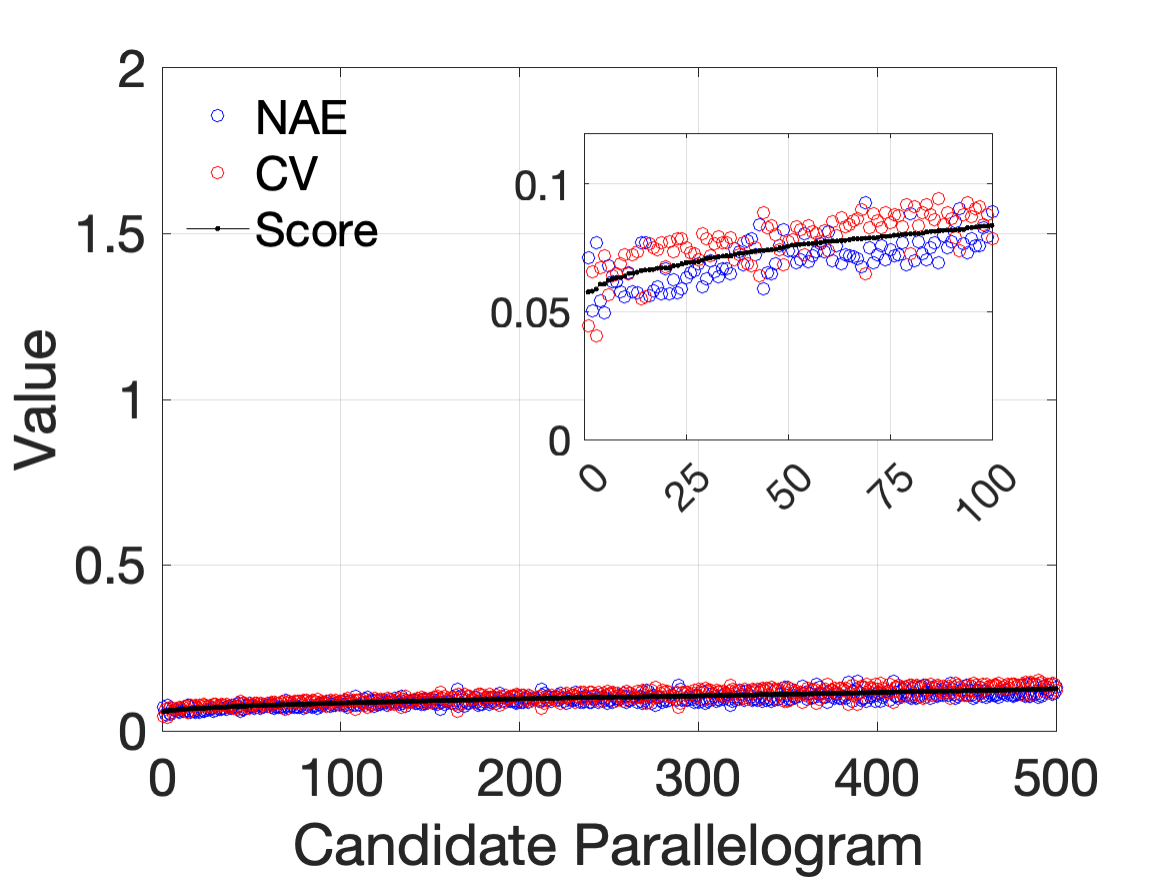}
        \caption{60 km/h}
    \end{subfigure}
    \caption{NGSIM: US101, Lane}
    \label{fig:parallelogram_metric:US101}
\end{figure}

%%%%%%%%%%%%%%%%%%
Figure~\ref{fig:TXPlot} visualizes the CV, NAE, and the scores.
In most cases with the given speed $v^*$ larger than 5 km/h, the CV and NAE show the same trend (positive corelation). 
When the given speed $v^*$ smaller than 5 km/h (Figure \ref{fig:parallelogram_metric:US101}(a)-(c)) are negatively correlated, which is mainly because the values are relatively more senstive when the speed is smaller.
The scores in general are smaller than 1 when $w_\text{CV} = 0.5$ and $w_\text{NAE}=0.5$. 
In particular, when the speed is high (such as 40 km/h and 60 km/h), the scores are quite small, indicating quite stationary traffic state.
%%%%%%%%%%%%%%%%%%

\begin{figure}[htbp]
    \centering
    \begin{subfigure}{0.32\textwidth}
    	\includegraphics[width=\linewidth]{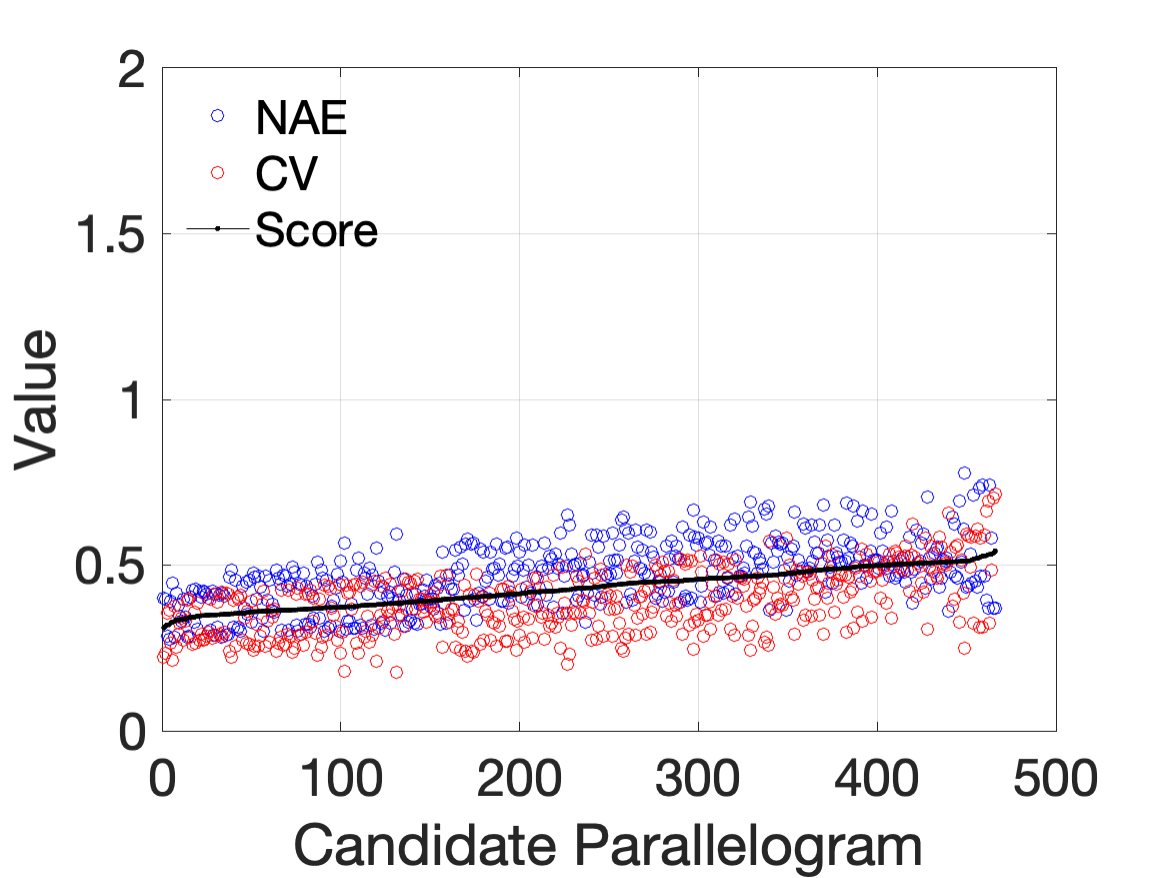}
        \caption{5 km/h}
    \end{subfigure}
    \begin{subfigure}{0.32\textwidth}
        \includegraphics[width=\linewidth]{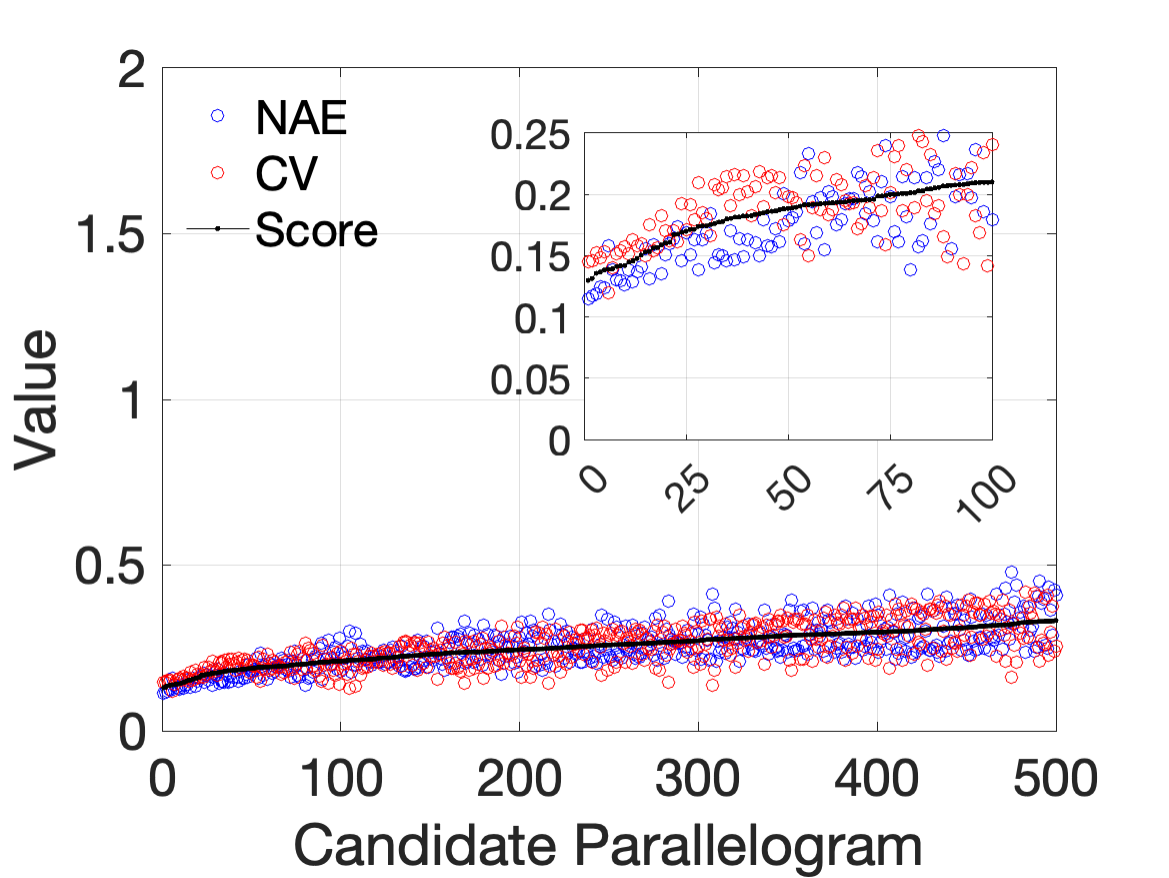}
        \caption{10 km/h}
    \end{subfigure}
    \begin{subfigure}{0.32\textwidth}
        \includegraphics[width=\linewidth]{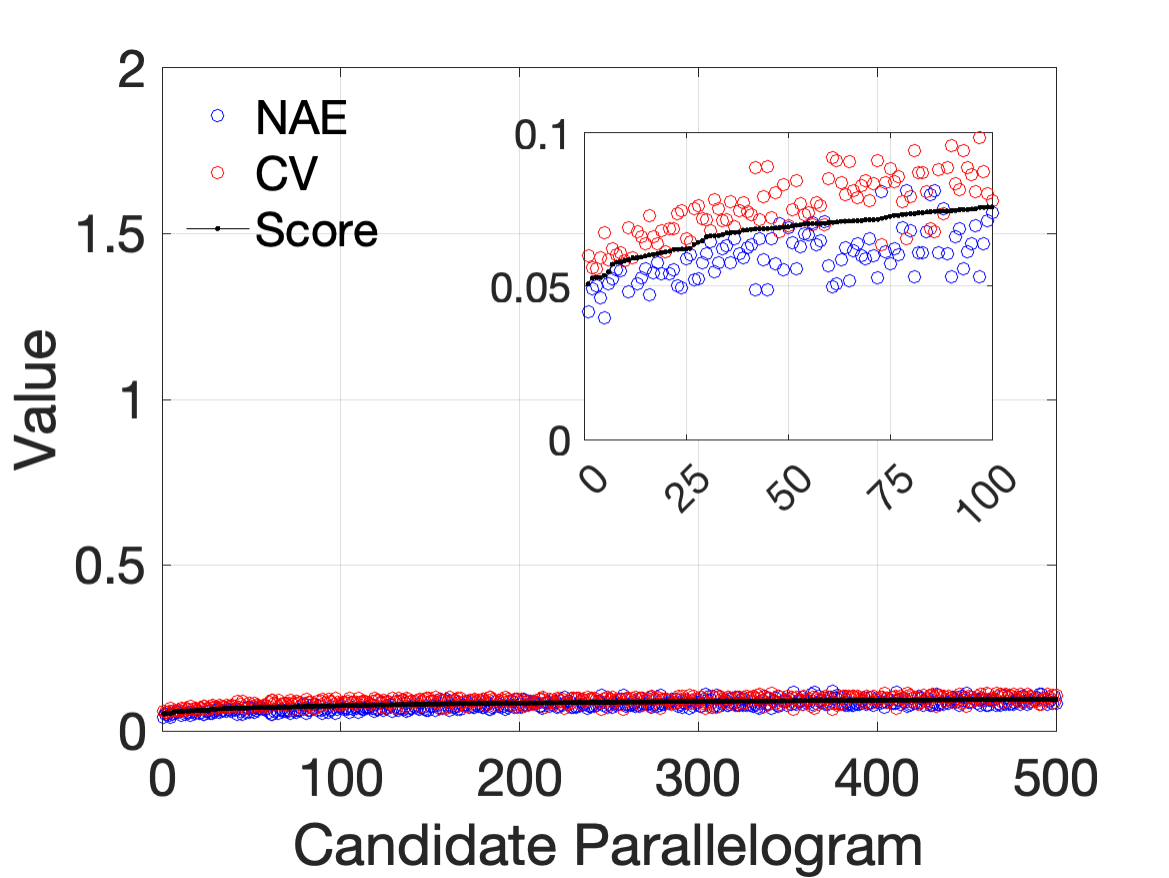}
        \caption{20 km/h}
    \end{subfigure}
    
    \begin{subfigure}{0.32\textwidth}
        \includegraphics[width=\linewidth]{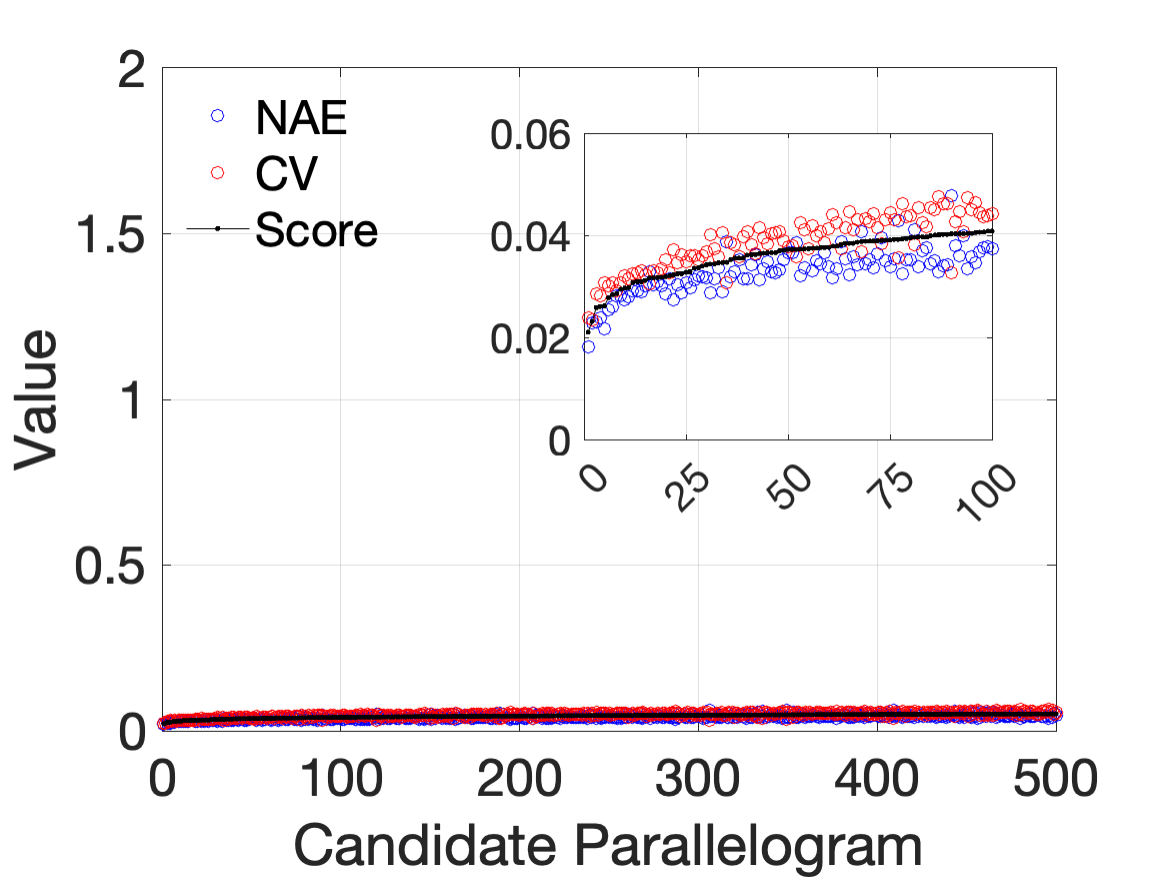}
        \caption{30 km/h}
    \end{subfigure}
    \begin{subfigure}{0.32\textwidth}
        \includegraphics[width=\linewidth]{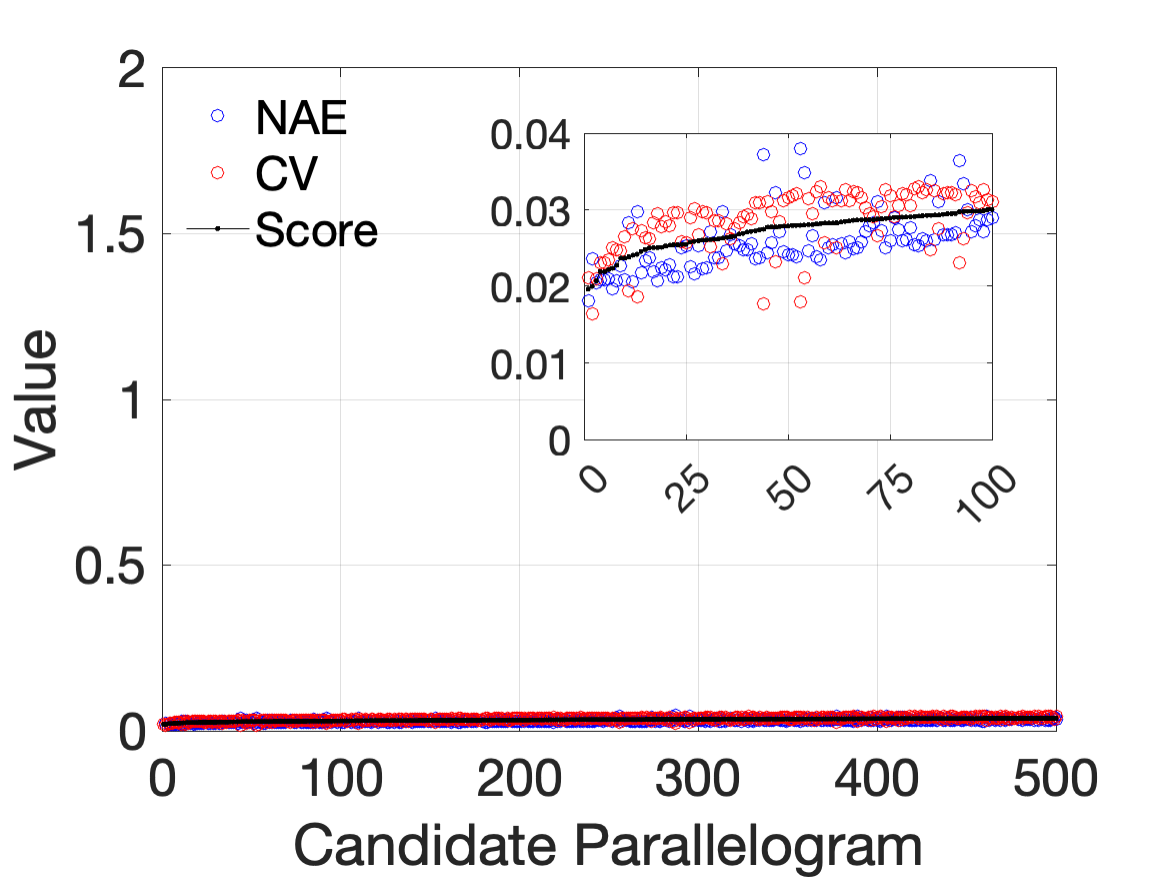}
        \caption{40 km/h}
    \end{subfigure}
    \begin{subfigure}{0.32\textwidth}
        \includegraphics[width=\linewidth]{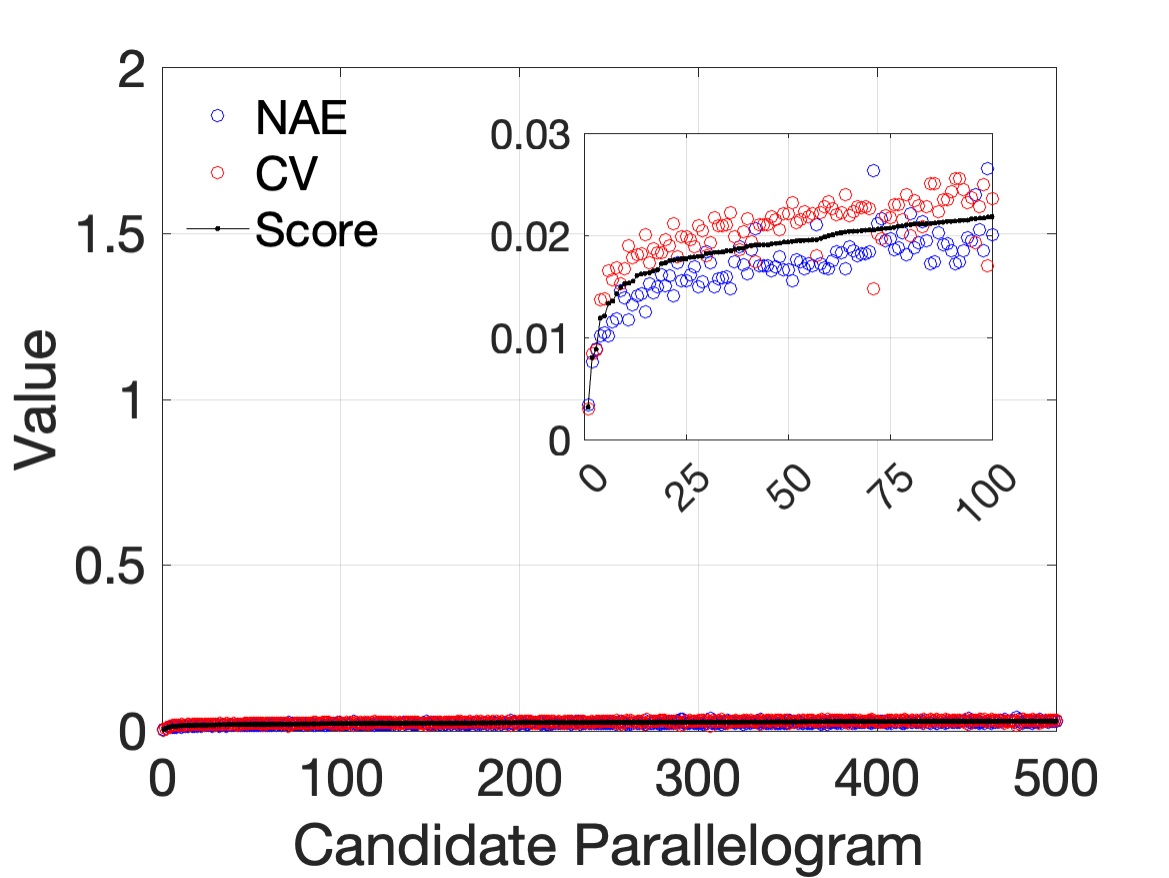}
        \caption{60 km/h}
    \end{subfigure}
    \caption{Zen Traffic: TRJ 11, Lane 1}
    \label{fig:parallelogram_metric:TRJ_11}
\end{figure}

\begin{figure}[htbp]
    \centering
    \begin{subfigure}{0.45\textwidth}
        \includegraphics[width=\linewidth]{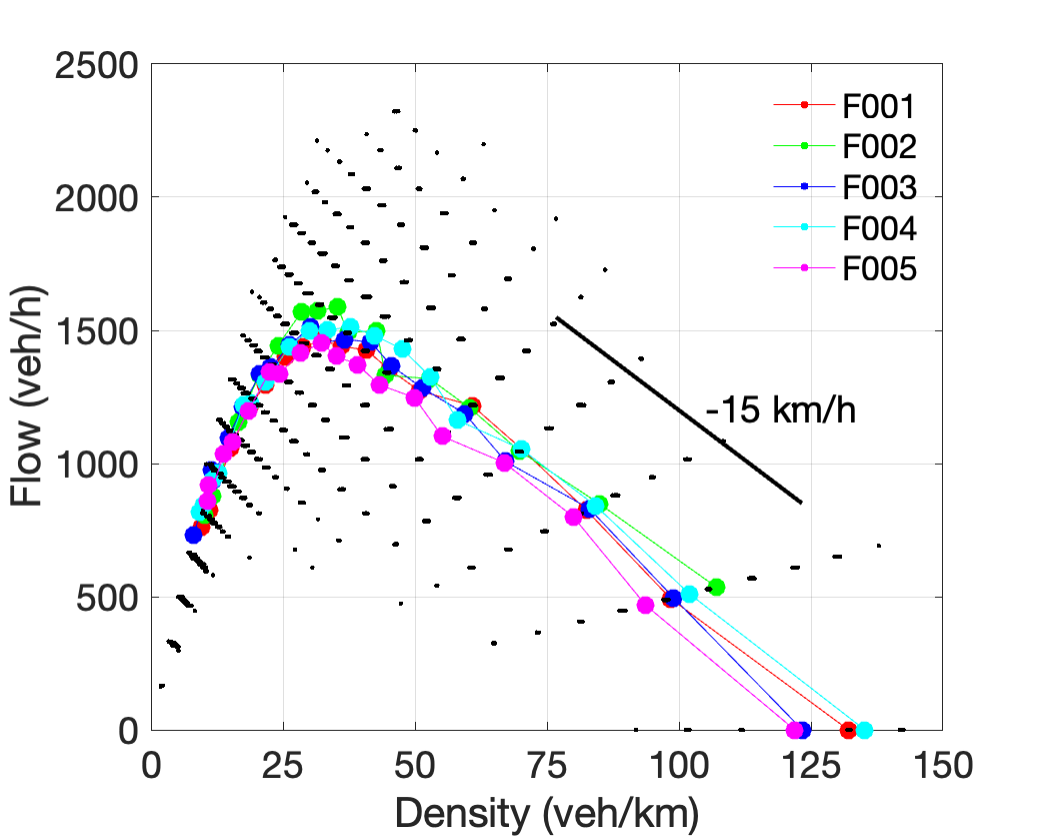}
        \caption{ZenTraffic (TRJ 11, Lane 1): Wave speed = -20 km/h}
    \end{subfigure}
    \begin{subfigure}{0.45\textwidth}
        \includegraphics[width=\linewidth]{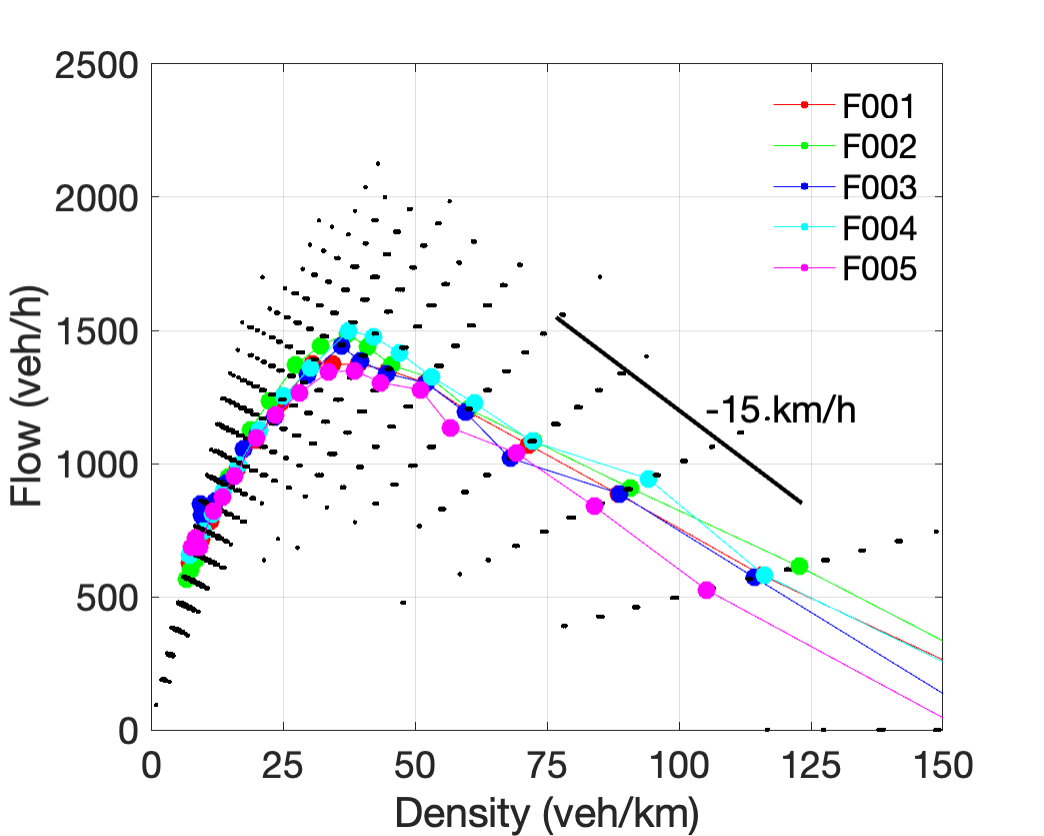}
        \caption{ZenTraffic (TRJ 11, Lane 1): Wave speed = -10 km/h}
    \end{subfigure}
    
    \begin{subfigure}{0.45\textwidth}
        \includegraphics[width=\linewidth]{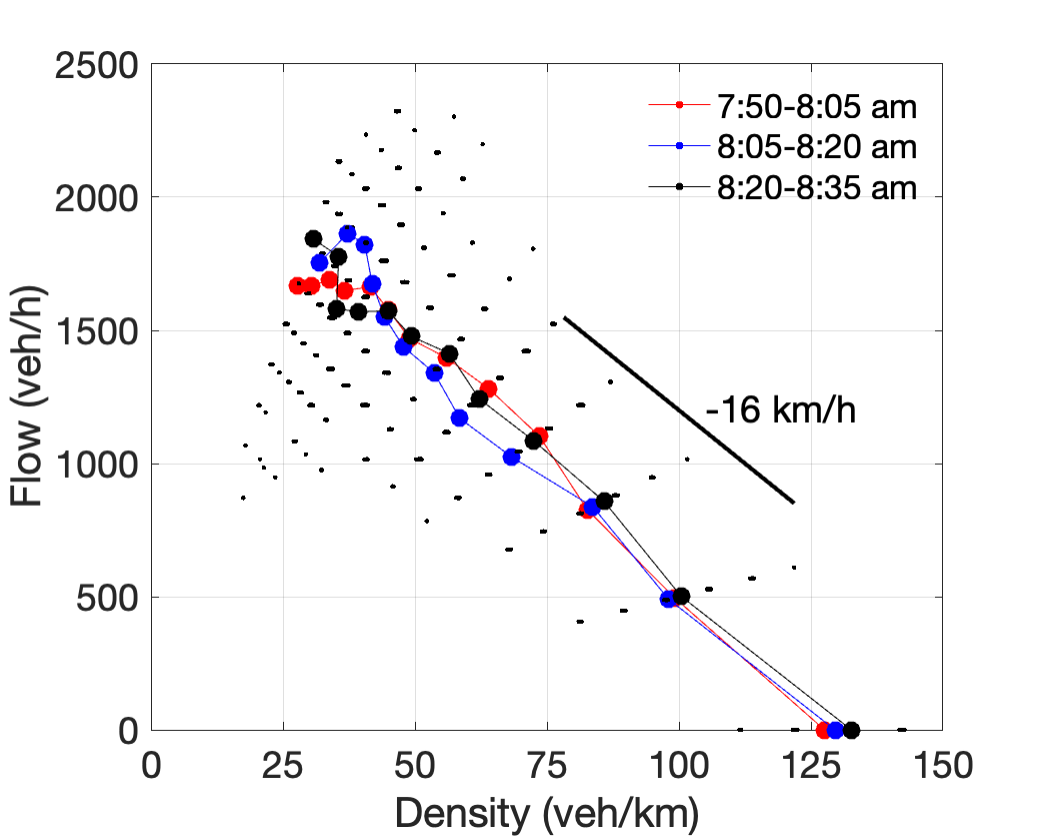}
        \caption{NGSIM (US101, Lane 1): Wave speed = 20 km/h}
    \end{subfigure}
    \begin{subfigure}{0.45\textwidth}
        \includegraphics[width=\linewidth]{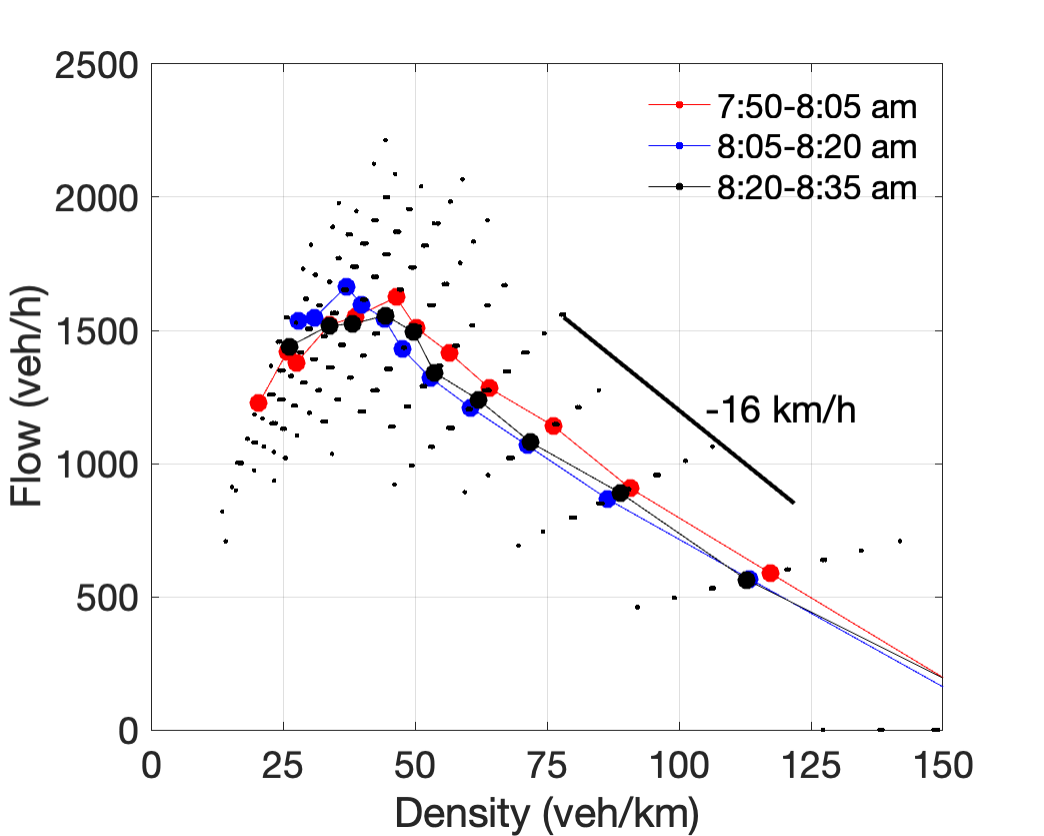}
        \caption{NGSIM (US101, Lane 1): Wave speed = -10 km/h}
    \end{subfigure}
    \caption{Empirical fundamental diagrams with different slopes of parallelograms.}
    \label{fig:Sensitivity}
\end{figure}

Wave speed is treated as an exogenous parameter in our method.
Although it generally lies within a narrow range (typically between -20 km/h and -10 km/h), conducting a sensitivity analysis helps assess its influence on the resulting fundamental diagram.
To this end, we re-ran the tool on the ZenTraffic and NGSIM datasets using two extreme wave speed values: -20 km/h and -10 km/h. The results are shown in Figure~\ref{fig:Sensitivity}.
When the wave speed is set to -20 km/h, the resulting congestion branches remain largely consistent with those in Figure~\ref{fig:FD}, aligning well with the observed wave propagation direction in the time-space diagram, as indicated by the black reference line in Figure~\ref{fig:Sensitivity}.
In contrast, using -10 km/h leads to a deviation: the resulting wave speed in the fundamental diagram is slower than the observed propagation speed in the time-space diagram.
This underscores the importance of carefully selecting the wave speed based on the empirical time-space diagram, where it can typically be identified through visual inspection.

\section{Future work}

Future research could focus on the following directions: (1) conducting systematic sensitivity analyses of critical parameters, such as parallelogram size and the weighting coefficients $w_\text{CV}$ and $w_\text{NAE}$, 
(2) investigating the causes of sensitivity to predefined wave speeds, and reducing the dependence on this parameter to enhance robustness, and (3) exploring new patterns of traffic flow and congestion with the improved fundamental diagram construction method proposed in this study.

%%%%%%%%%%%%%%%%%%%%%%%%%%%%%%%%%%%%%%%%%%%%%%%%%%%%%%%%%%%%%%%%%%%%%%%%%%%%%%%%%%%%%%
%%%%%%%%%%%%%%%%%%%%%%%%%%%%%%%%%%%%%%%%%%%%%%%%%%%%%%%%%%%%%%%%%%%%%%%%%%%%%%%%%%%%%%
%\section*{Acknowledgement}

%The research is funded by 
%National Key R\&D Program of China (2018YFB1600505).
%National Natural Science Foundation of China (71871010).
%National Key R\&D Program of China (2018YFB1601302).


\bibliographystyle{model2-names}
\bibliography{library}

\small
\section*{Afterword}
\textit{Motivated by a personal commitment to the topic, the first author has pursued this line of research despite limited funding support. However, due to resource constraints, he has been unable to dedicate the time required for a more in-depth investigation, even though there remains substantial detailed work to be done. The first author hopes that future opportunities will allow him to complete this research in a more comprehensive manner.}

\end{spacing}
\end{document}